\documentclass[review]{elsarticle}

\usepackage{tikz}
\usepackage{amsmath}
\usepackage{amssymb}
\usepackage{booktabs}
\usepackage{soul}
\usepackage{subfig}
\usepackage{multirow}
\usepackage[export]{adjustbox}
\usepackage{ulem}
\usepackage{algorithm}
\usepackage{algpseudocode}

\usetikzlibrary{shapes.geometric,backgrounds, arrows,calc,fit}
\tikzstyle{large_block} = [rectangle, rounded corners, minimum width=2cm, text width=5.6cm, minimum height=1cm,text centered, draw=black]
\tikzstyle{super_large_block} = [rectangle, rounded corners, minimum width=2cm, text width=7.0cm, minimum height=1cm,text centered, draw=black]
\tikzstyle{arrow} = [thick,->,>=stealth]

\tikzset{
  basic box/.style = {
    shape = rectangle,
    align = center,
    draw  = #1,
    fill  = #1!5,
    rounded corners}
}
\def\connectwithlongtext#1#2#3#4{ 
\draw[#1] 
  let
    \p1 = ($(#2)-(#3)$),
    \n1 = {0.6*veclen(\x1,\y1)}
  in  (#2) -- (#3)
   node[midway, below, align=center, text width=\n1, sloped] 
   {#4};
}

\begin{document}

\begin{frontmatter}

\title{Optimised graded metamaterials for mechanical energy confinement and amplification via reinforcement learning$^\star$}
\tnotetext[t1]{This is the accepted version of \textit{L. Rosafalco, J. M. De Ponti, L. Iorio, R. Ardito, A. Corigliano, Optimised graded metamaterials for mechanical energy confinement and amplification via reinforcement learning, Eur. J. Mech. A Solids, 104947 (2023)}. The final publication is available at \texttt{https://doi.org/10.1016/j.euromechsol.2023.104947}.}

\author[1]{Luca Rosafalco}
\ead{luca.rosafalco@polimi.it}
\author[1]{Jacopo Maria De Ponti}
\ead{jacopomaria.deponti@polimi.it}
\author[1]{Luca Iorio}
\ead{luca.iorio@polimi.it}
\author[1]{Raffaele Ardito}
\ead{raffaele.ardito@polimi.it}
\author[1]{Alberto Corigliano}
\ead{alberto.corigliano@polimi.it}
\address[1]{Dipartimento di Ingegneria Civile ed Ambientale,
Politecnico di Milano \\ Piazza L. da Vinci 32,
20133 - Milano (Italy)}

\begin{abstract}
A reinforcement learning approach to design optimised graded metamaterials for mechanical energy confinement and amplification is described. Through the proximal policy optimisation algorithm, the reinforcement agent is trained to optimally set the lengths and the spacing of an array of resonators. The design optimisation problem is formalised in a Markov decision problem by splitting the optimisation procedure into a discrete number of decisions. Being the physics of graded metamaterials governed by the spatial distribution of local resonances, the space of possible configurations is constrained by using a continuous function for the resonators arrangement. A preliminary analytical investigation has been performed to characterise the dispersive properties of the analysed system by treating it as a locally resonant system. The outcomes of the optimisation procedure confirms the results of previous investigations, highlighting both the validity of the proposed approach and the robustness of the systems of graded resonators when employed for mechanical energy confinement and amplification. The role of the resonator spacing is shown to be secondary with respect to the resonator lengths or, in other words, with respect to the oscillation frequencies of the resonators. However, it is also demonstrated that reducing the number of resonators can be advantageous. The outcomes related to the joint optimisation of the resonator lengths and spacing, thanks also to the adaptive control of the analysis duration, overcome significantly the performance of previously known systems by working almost uniquely on enlarging the time in which the harvester oscillations take place without amplifying these oscillations. The proposed procedure is suitable to be applied to a wide range of design optimisation problems in which the effect of the design choices can be assessed through numerical simulations.

\end{abstract}

\end{frontmatter}


\section{Introduction}
\label{sec:introduction}

In the last two decades metamaterial concepts have witnessed an increasing popularity to control the propagation of waves across much of physics and engineering, with multiple realizations in electromagnetism \cite{art:Smith2004,art:Pendry1999,art:Pendry2000}, acoustics \cite{art:Liu2000, book:Craster2013} and elasticity \cite{book:Craster2017}. In the context of elasticity, considerable effort has been devoted to the investigation of novel mechanisms to manipulate elastic waves for numerous applications of technological relevance such as, nondestructive evaluation \cite{art:Moleron2015,art:Ali2021}, vibration isolation \cite{book:Laude2015,Matlack2016}, seismic protection \cite{Brule2014,Miniaci2016,Brule2020} and cloaking \cite{Farhat2009,Stenger2012, Quadrelli2021} to name a few. Concurrently, optimal design solutions have been developed \cite{KRUSHYNSKA2014179}, together with advanced modelling methods  for complex metamaterials \cite{MATOUS2017192,Sridhar2016,LIU2021114161}. Metamaterials are often combined with multi-physics materials, leveraging energy conversion phenomena between mechanical deformations and, for instance, electrical stimuli via piezoelectric coupling. In this context, multifunctional metamaterials have been recently proposed for energy harvesting purposes, thanks to their ability to simultaneously provide vibration isolation and mechanical energy enhancement \cite{Carrara2013, Mikoshiba2013, Gonella2009, Sugino2018, Chen2019}. Whilst local resonators in elastic metamaterials allows for strong energy enhancement, the harvested power is noticeable only close
to the bandgap frequency and energy confinement is accompanied by strong scattering effects \cite{chp:DePonti2021}.


To overcome such limitation, a versatile way to obtain inherent broadband and low scattering designs is based on graded metamaterials, which are structures incorporating the gentle variation of resonating elements. The term \textit{graded} refers to a smooth variation of a particular parameter of the local resonators along space (conventionally the resonance frequency), which enables spatially varying effective properties of the medium. These devices take advantage of local band gaps to control wave propagation; array guided waves slow down as they transverse the array with different frequency components localising at specific spatial positions, resulting in the so-called rainbow effect. Originated in electromagnetism using axially non-uniform, linearly tapered, planar waveguides with cores of negative index material \cite{art:Tsakmakidis2007}, the rainbow effect has been extended to acoustics \cite{Zhu2013, Garcia2013, Garcia2014} and elasticity \cite{art:Colombi2016}, with multiple realizations for trapping \cite{Chaplain2020Delineating} or mode conversion \cite{art:Chaplain2020Umklapp}. Graded metamaterials have thus attracted increasing attention due to their ability to manipulate waves by confinement over some spatial region along the structure, enabling wideband vibration attenuation. Within this framework, a number of graded metamaterials have been proposed for wave confinement and energy harvesting, using elastic beams with graded resonators \cite{art:NJP20,art:APL20, Chaplain2020Delineating, Zhao2022}.

Even if several works have been done on graded arrays, the definition of an optimal spatial modulation of the medium properties for energy confinement and amplification is still an unsolved problem. In acoustics, a stronger sound enhancement has been demonstrated in exponentially chirped crystals rather than in linearly chirped crystals \cite{Garcia2014}. In elasticity, several grading laws have been compared, and different performances both in terms of vibration isolation and energy harvesting have been reported \cite{ErturkGrading2020,ErturkGrading2022,ErturkGrading2022b,JianGrading2022,Zhao2022}. 
In this article, a general procedure for the grading optimisation (both in terms of frequency and spacing) is proposed. In doing so, we opt for an elastic beam with lumped resonators, as commonly done in the literature on graded metamaterials. This allows us to define a relatively simple 1D proof of concept framework, which could be further generalised to 2D and 3D. Specifically, the structure under study is a rainbow based metamaterial consisting of a graded array of resonant rods connected to an elastic waveguide \cite{art:NJP20,art:APL20}. As the resonance frequencies are primarily determined by the resonator lengths, we have treated the resonator lengths and spacing as the main design parameters.
Moreover, for the sake of simplicity, we focus on the mechanical problem only, neglecting the conventional transduction mechanisms adopted for energy harvesting; this allows us to identify optimal solutions for energy confinement, without loosing generality about the wave propagation problem.\\

A Markov Decision Process (MDP) has been used to formalise the optimal design search as done in \cite{art:Ororbia21}. Reinforcement Learning (RL) \cite{book:Sutton18} has been then employed to solve the MDP. The use of RL and of the MDP formalisation concepts have been preferred: (i) to gradient based methods, as these approaches would have been negatively affected by the non regularity of the optimisation task with respect to the design parameters \cite{SKINNER2018933}, e.g. the number of resonators; (ii) to genetic algorithms \cite{art:Jenkins91}, as they suffer from an high computational cost; (iii) to particle swarm optimisation \cite{art:Perez07}, due to the difficulty of imposing constrains on the design parameters \cite{art:Viquerat21}. Moreover, RL enjoys the theoretical advantage of handling possible source of stochasticity affecting the optimal design search. It is our aim to investigate this aspect in future work.

In order to provide information to the RL agent, the Finite Element Method (FEM) has been employed on an Euler-Bernoulli beam with lumped resonators to evaluate the response of different arrangements of resonators. In \cite{art:Fan20}, a RL agent exploited experimental data to discover active control strategies for drag reduction and to gain an insight on the problem, showing that best design solutions often trigger particular physical phenomena. Similarly, in \cite{art:Pahlavani22} rare--event designs of 3D printed multi--material metamaterials were found out by learning a map from the space of design parameters to the space of mechanical properties through deep learning.

Another aspect of interest is the representation strategy handling the space of the design parameters. Similarly to what was done by \cite{art:Papadrakakis98} in structural shape optimisation, the possible resonator arrangements have been described through the coordinates of few points by exploiting B--spline interpolation. In this way, a large number of configurations and possible modifications of the resonator arrangement could been obtained by playing on a limited number of variables. This approach has been suggested by the physics of the problem. Paying attention to the representation strategy is one of the points of contact between the proposed approach and a family of strategies for automatic system design termed computational design synthesis \cite{art:Cagan05}. A second even more evident similarity is the strong automation of the design process.

In the light of the above, the main contribution of this work is the combination of: the MDP formalisation of the design optimisation problem; the constraining of the design space based on physical consideration; the use of an actor--critic RL algorithm for solving the MDP. The potentiality of the method is demonstrated by presenting optimised configurations for a 1D rainbow based metamaterial in which variable resonator lengths or variable resonator spacing, or variable resonator lengths and spacing have been respectively considered. Interestingly, a different number of resonator bars has resulted from the performed optimisations, showing that employing the maximum allowed number of resonators can be useless or even detrimental. Obtained configurations can be interpreted in a physical sense in accordance with previous works \cite{art:NJP20,art:APL20}, but revealing aspects of novelty. As it is in the following be shown, the best grading rule privileges the extension of the 
target resonator oscillation time with respect to the amplification of the wave excitation along the guide. The methodology can be straightforwardly applied to other design optimisation problems in which design choices can be evaluated through numerical simulations. A certain robustness to the algorithm hyperparameters, here almost identically employed in the two optimisation cases, further increases the attractiveness of the method.

The reminder of the paper is arranged as follows. The proposed approach is illustrated in Sec. \ref{sec:methodology}, together with an introduction to MDP and RL. The discussion and motivation of the representation describing the possible design arrangements, exploiting few points and B--spline interpolation, is also included. In Sec. \ref{sec:results}, the optimisation outcomes concerning the cases featuring variable resonator lengths (Sec. \ref{sec:resultsLengths}), and variable resonator lengths and spacing (Sec. \ref{sec:resultsLengthSpacing}) are presented, together with physics based interpretations justifying their improved performance. These results are preceded in Sec. \ref{sec:dispersionRelation} by an insight on the mechanical problem obtained by working out the dispersion relation for the rainbow based metamaterial when represented as a waveguide with resonators. Final considerations on the proposed methodology and on the obtained resonator configurations are contained in Sec. \ref{sec:conclusions}, together with a discussion on the future developments of this work.

\section{Methodology}
\label{sec:methodology}

\begin{figure}[H]
\begin{centering}
\begin{tikzpicture}[node distance=2cm]
\node (define_config) at (0, 5) [large_block]  {Define the space \\ of possible configurations.};
\node (define_goal) [large_block, right of=define_config, xshift=4cm] {Define the \\ optimisation goal.};
\begin{scope}[on background layer]
\node[fit = (define_config)(define_goal), basic box = white] (define) {};
\end{scope}
\node (constrain_design) [super_large_block] at (3, 3) {Constrain the design space \\ exploiting physical knowledge.};
\node (split_design) [super_large_block] at (3, 1) {Split the design process \\ into a sequence of decisions.};
\node (formalization) [super_large_block] at (3, -1) {Formalise the design process \\ as a Markov Decision Process.};
\node (solve) [super_large_block] at (3, -3) {Solve the Markov Decision Process \\ via Reinforcement Learning.};
\connectwithlongtext{->}{define}{constrain_design}{ };
\connectwithlongtext{->}{constrain_design}{split_design}{ };
\connectwithlongtext{->}{split_design}{formalization}{ };
\connectwithlongtext{->}{formalization}{solve}{ };
\end{tikzpicture}
\caption{{Methodology to design optimisation.}\label{fig:MethodologyGeneral}}
\end{centering}
\end{figure}
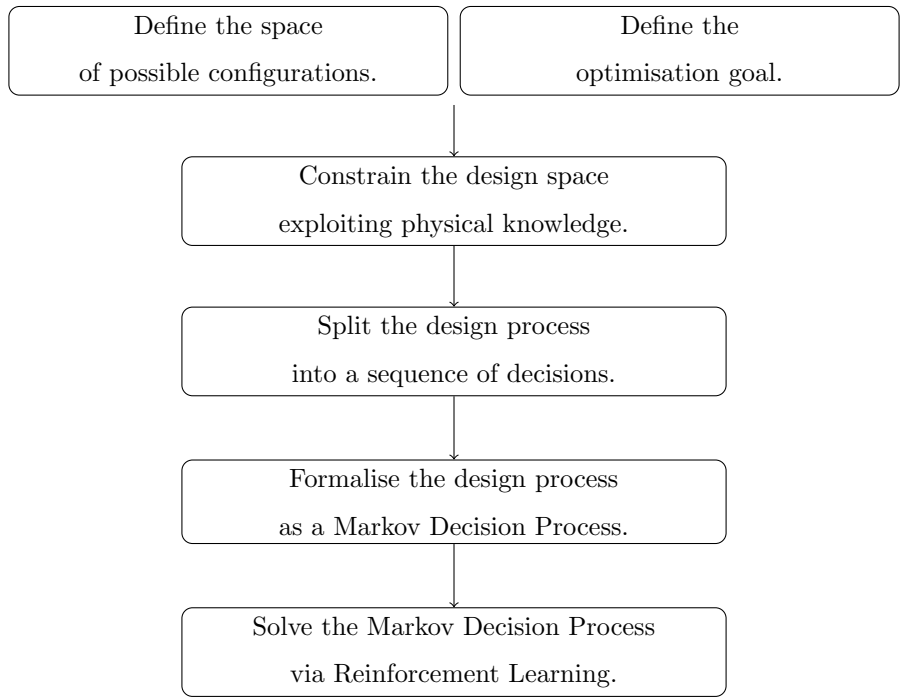

\begin{figure*}
\centering
\begin{tikzpicture}
  \node (geom)  {\includegraphics[width=110mm]{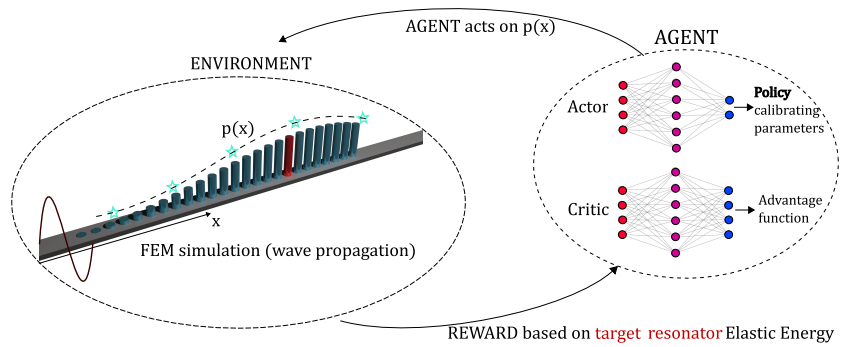}};
 \end{tikzpicture}
\caption{\footnotesize Reinforcement learning for grading optimisation for mechanical energy confinement and amplification. RL exploits information coming from the interaction between the environment (on left) and the agent (on right). The environment is defined by the wave propagation problem. The RL agent acts on the grading rule $p\left(x\right)$ of the rainbow based metamaterial modifying few interpolation points (cyan markers) according to a certain policy. The policy is modelled through a neural network receiving as inputs the coordinates of the interpolation points. The critic provides information to modify the policy on the basis of the agent--environment interactions. In particular, finite element simulations are employed to assess possible enhancements of the mechanical energy confinement for a modified grading rule. \label{fig:Methodology}}
\end{figure*}

The operation workflow for design optimisation is reported in Fig. \ref{fig:MethodologyGeneral}. While the need of constraining the design space will be addressed in the following, it is considered that if no a priori knowledge of the final design is available, the optimisation of a complex mechanical system is usually split into a sequence of $N_t$ decisions, producing a set of $N_t$ configurations defining specific design descriptions of the systems. In the following, decisions are termed actions, and configurations are named states, while a sequence of actions and configurations leading to a final design is defined trajectory. In particular, the $n_t$th configuration $S_{n_t}$, together with the corresponding reward $R_{n_t}$, is obtained by modifying $S_{n_{t-1}}$ through the $t$th action $A_{n_t}$. The choice of the action $A_{n_t}$ is based on the information collected into $S_{n_{t-1}}$. The sequential decision making process can be formalised in a MDP as the probability to end in $S_{n_t}$ depends only on $S_{n_{t-1}}$ and on $A_{n_{t-1}}$. Solving a MDP via RL requires to define two entities interacting with each other, namely the agent and the environment. Actions are said to be taken by an agent, while states and rewards are said to be related to an environment. For the case at hand, a sketch of this interaction is reported in Fig. \ref{fig:Methodology}. A reward measures how well the optimisation task is pursued in a certain state. Here, the state of the mechanical system is defined by the resonator arrangements, while actions modify the resonator lengths and/or spacing. The reward is defined by the sum over the analysis duration of the harvested elastic energy $\mathcal{E}$, as enhancing $\mathcal{E}$ is hypothesised to be related to the confinement and amplification of the wave packet.

It is now useful to introduce some quantities that will be exploited by the employed actor--critic RL algorithm. First, we observe that the interest is not to maximise the immediate reward $R_{n_t}$ at a generic $n_t<N_t$, but to set a strategy to achieve large rewards at the end of the optimisation process for $n_t=N_t$. For this reason, the notion of expected return is employed. It consists in evaluating how good is to end up in $S_{n_t}$ after $n_t$ actions and then adopting a certain strategy to take the subsequent decisions. For a finite MDP featuring $N_t<\infty$, the expected return $G_{n_t}$ is usually defined as:

\begin{equation}
    G_{n_t} = R_{n_{t+1}}+R_{n_{t+2}}+...+R_{N_t} .
    \label{eq:expectedReturn}
\end{equation}

Second, the concept of policy is introduced. The policy $\pi$ defines the strategy guiding the agent decisions. If a deterministic policy $\pi$ is employed, a unique action is associated to each state; if $\pi$ is stochastic, a Probability Density Function (PDF) over the set of possible actions is returned. An example of PDF is the one related to a Gaussian distribution. A suitable policy allows to solve the MDP by setting a certain sequence of actions. Here, the wanted policy is deterministic, because no uncertainties affect the design at the considered stage. However, during the optimisation it is useful to employ stochastic policies to guarantee the exploration of the action and state spaces.
 
Last, the concept of value function is discussed. The value function $v_{\pi}\left(s\right)$ of a state $s$ (here intended as a random variable, while $S_{n_t}$ indicates a possible realisation of $s$ at $n_t$) under a policy $\pi$ is defined by exploiting the notion of expected return as

\begin{equation}
    v_{\pi}\left(s\right) = \mathbb{E}_{\pi}\left[G_{n_t}|S_{n_t}=s \right],
    \label{eq:valueFunction}
\end{equation}

where $\mathbb{E}_{\pi}$ denotes the expected value under $\pi$ or, in other words, the expected value computed starting from $s$ and following $\pi$ thereafter.

By considering as input space the one defined by the combination of the state and action spaces, two functions, namely the action--value function $q_{\pi}\left(s,a\right)$ and the advantage function $d_{\pi}\left(s,a\right)$ connected to $v_{\pi}\left(s\right)$ are introduced

\begin{subequations}
\begin{gather}
    \label{eq:actionValueFunction}
    q_{\pi}\left(s,a\right) = \mathbb{E}_{\pi}\left[G_{n_t}|S_{n_t}=s,A_{n_t}=a \right], \\
    d_{\pi}\left(s,a\right) = q_{\pi}\left(s,a\right) - v_{\pi}\left(s\right),
\end{gather}
\end{subequations}

where $a$ is an action, here intended as random variable, while $A_{n_t}$ is its possible realisation at $n_t$. Specifically, $q_{\pi}\left(s,a\right)$ quantifies the expected return of $s$ when $a$ is taken and then $\pi$ is adopted; $d_{\pi}\left(s,a\right)$ is the advantage function. Within a MDP, treating $d_{\pi}\left(s,a\right)$ is often preferred than handling $v_{\pi}\left(s\right)$ because the estimation of the advantage function is less affected by variance than the estimation of the value function \cite{art:Greensmith04}.

The concept of value function is important because by evaluating if $v_{\pi}\left(s\right)\geq v_{\bar{\pi}}\left(s\right)$ in every state, it is possible to conclude that a policy $\pi$ is better or equal than a policy $\bar{\pi}$. The policy gradient method aims to learn the best policy $\pi^{\ast}\left(a|s\right)$ \cite{art:Barto83}. In RL, $\pi^{\ast}$ is searched by exploiting indications coming from a large number of agent--environment interactions. In contrast, action--value methods do not explicitly look for the best policy focusing, first, on estimating $v\left(s\right)$ and, secondly, in reconstructing the best policy by picking up the states with the largest value functions.

In this work, policy gradient methods have been preferred \cite{book:Sutton18}. First, they allow for asymptotically approaching a deterministic policy starting from a stochastic policy. Secondly, they have the capacity of automatically learning appropriate levels of exploration. Moreover, they enjoy the theoretical advantage of expressing the effect of a policy change on the value functions without computing derivatives with respect to the state distribution. This is guaranteed by the policy gradient theorem \cite{proc:Sutton99}, which is at the basis of the employed RL algorithm. Last, policy gradient methods easily handle continuous action spaces.

The convenience of constraining the design space exploiting physical knowledge is now discussed. If a numeric vector collecting the resonator lengths is adopted to represent the state, the Markovianity of the sequential decision process will be fulfilled as all the information necessary to define the reward and to plan the next action will be contained in the state representation. If we set as action the modification of a single resonator height, and if we fix the number of resonators and the number of possible levels for the resonator lengths, a discrete number of states will be treated. Discretising the resonator length will not limit the number of obtainable configuration because, on one hand, the discretisation level can be arbitrary increased, while on the other hand the tolerances always present in a manufacturing process agrees with this schematisation. This state definition promises to easily solve the optimisation process, e.g. through a brute force approach consisting in investigating all the states of the system. However, proceeding in this way is generally impossible for the exploding number of combinations. Things do not improve by adopting a RL approach, given that modifying the resonator height one by one does not produce large changes in the rewards, providing weak indications to the agent on how to improve the starting configuration. In other words, if we look at the tackled optimisation task as a gradient based function optimisation, agent explorations corresponds to function evaluations in flat regions.

\begin{figure*}
\centering
\begin{tikzpicture}
  \node (geom)  {\includegraphics[width=100mm]{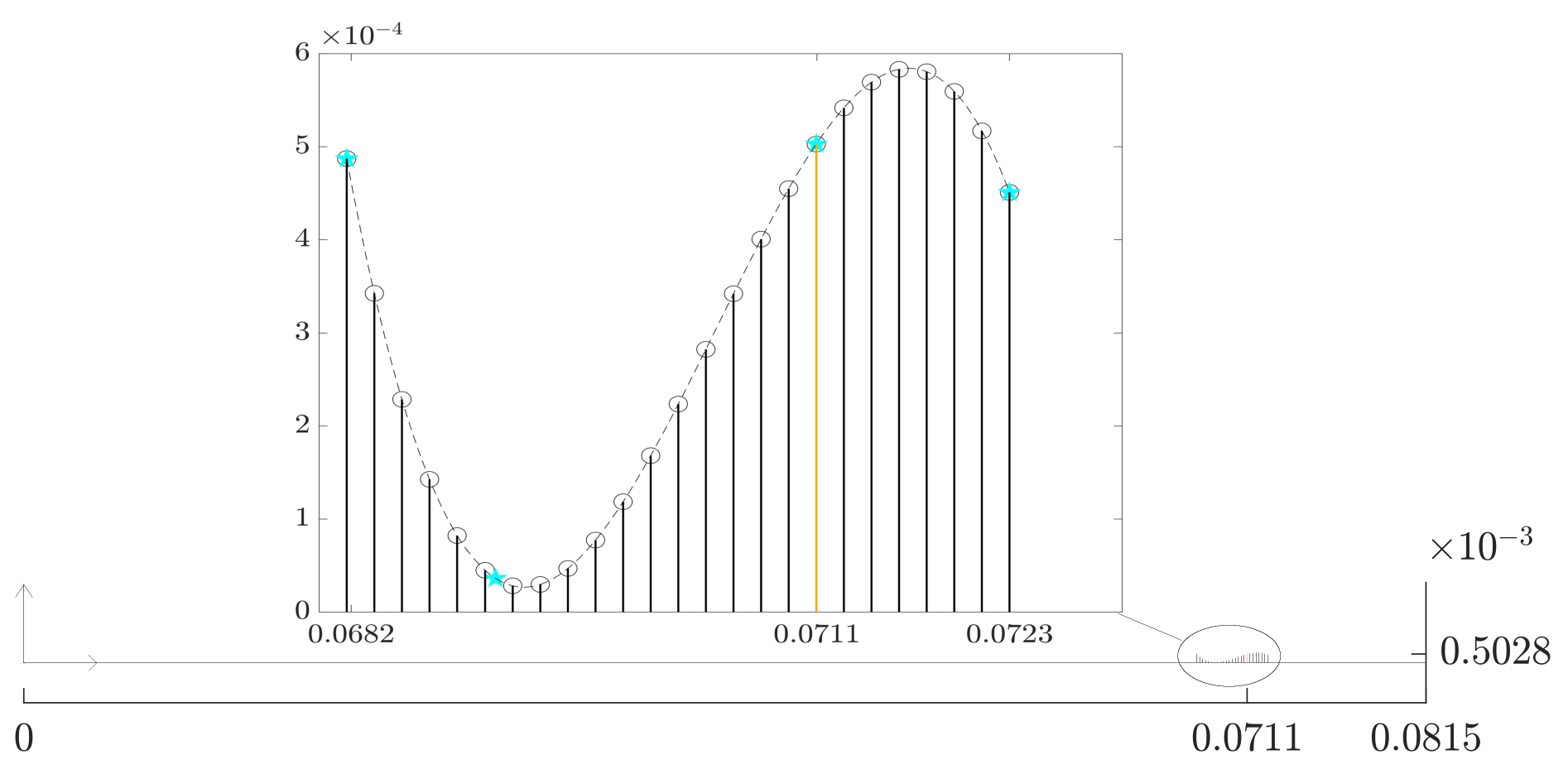}};
  \node[below of=geom,node distance=0cm,xshift=-5.3cm, yshift=-1.4cm,font=\color{black}] {\footnotesize $z\text{}\left[\text{m}\right]$};
  \node[below of= geom,node distance=0cm,xshift=-4.5cm, yshift=-1.55cm,font=\color{black}] {\footnotesize $x\text{}\left[\text{m}\right]$};
 \end{tikzpicture}
\caption{\footnotesize Reconstruction of the resonator lengths starting from interpolation points (depicted by cyan markers) whose positions are set by the state variables. The envelope curve $z=p\left(x\right)$ is depicted by a dotted line. The harvester is plotted in orange, the other resonators in black. The circle at the resonator end recalls the mass--spring schematisation employed in the numerical modellisation of the system.\label{fig:interpolationExample}}
\end{figure*}

A more convenient definition of the system state has come from adopting a limited number of continuous variables $N_s$ to describe the possible resonator arrangements. This has been made possible by constraining the design space including the most physically meaningful configurations. From a ML perspective, the variance of the optimisation process has been reduced by introducing a bias based on a previously formed understanding of the problem. The adopted constrain consists in enforcing smooth graded patterns for the resonator lengths in agreement both with theoretical \cite{art:NJP20} and experimental \cite{art:APL20} results. The state continuous variables are used to fix the position of a few interpolation points defining the envelope  $z=p\left(x\right)$ of the resonator lengths, as shown in Fig. \ref{fig:interpolationExample}. Cubic B--splines are exploited for the interpolation. The agent modifies the coordinates of the interpolation points. The sequence of actions is fixed: in the $t$th episode of the agent--environment interaction, the agent always sets the position of a certain interpolation point. Thanks to the concept of expected return, defining a priori an order of operation does not affect the optimisation outcome. The effectiveness of the proposed state and action definitions will be further assessed by showing that increasing the number of state variables does not necessary benefit the optimisation procedure.

To treat continuous policy and action spaces, the state value functions have been approximated by parameterised functions as

\begin{subequations}
\begin{gather}
    v_{\pi}\left(s\right) \approx v\left(s,\boldsymbol{\theta}_v\right), \\
    q_{\pi}\left(s,a\right) \approx q\left(s,a,\boldsymbol{\theta}_v\right),
\end{gather}
\label{eq:valueFunctionsApproximation}
\end{subequations}
$\!$where $\boldsymbol{\theta}_v\in\mathbb{R}^{N_{\theta v}}$ is a vector of tunable weights.

Similarly, associating a PDF featuring a Gaussian distribution to the policy, deterministic function approximators have been used to parametrise the dependence on the state of the policy mean and standard deviation as

\begin{equation}
    \pi\left(a|s\right) = \frac{1}{\sqrt{2\pi}\sigma\left(s,\boldsymbol{\theta}_p\right)}\text{e}^{-\frac{1}{2}\left(\frac{a-\mu\left(s,\boldsymbol{\theta}_p\right)}{\sigma\left(s,\boldsymbol{\theta}_p\right)}\right)^2},
    \label{eq:parametricAppeox}
\end{equation}
where here $\pi$ is just the number $\approx 3.1415\ldots$; $\mu$ and $\sigma$ are two function approximators relying on the parameter vector $\boldsymbol{\theta}_p\in\mathbb{R}^{N_{\theta p}}$.

The way in which the MDP is solved by RL is now explicitly addressed. In particular, the Proximal Policy Optimisation (PPO) algorithm \cite{art:Schulman17} has been employed to approximate the optimal $\pi^{\ast}$ finally solving to the MDP by tuning $\boldsymbol{\theta}_p\in\mathbb{R}^{N_{p}}$. The PPO algorithm belongs to a subset of policy gradient methods named actor--critic approaches. In these methods, state value functions ($d_{\pi}$ for PPO) are used to assign credit to the agent actions. This requires the algorithm both to approximate the advantage function $d_{\pi}\left(s,a\right)\approx d\left(s,a,\boldsymbol{\theta}_{v}\right)$ and to calibrate the policy parameters. Two fully connected Neural Networks (NNs), ruled respectively by $\boldsymbol{\theta}_v$ and $\boldsymbol{\theta}_p$, have been used for this goal. The NN differentiability has been exploited within the backpropagation algorithm \cite{art:Rumelhart86} to maximise the PPO objective

\begin{equation}
\begin{split}
    &\mathcal{L}_p\left(\boldsymbol{\theta}_{p}\right)=\hat{\mathbb{E}}_{e}\Biggl[\text{min}\Biggl(\frac{\pi\left(a|s,\boldsymbol{\theta}_{p}\right)}{\pi_{\text{old}}\left(a|s,\boldsymbol{\theta}_{p_{\text{old}}}\right)}d\left(s,a,\boldsymbol{\theta}_v\right),\\
    & \quad\text{clip}\left(\frac{\pi\left(a|s,\boldsymbol{\theta}_{p}\right)}{\pi_{\text{old}}\left(a|s,\boldsymbol{\theta}_{p_{\text{old}}}\right)},1-\epsilon,1+\epsilon\right)d\left(s,a,\boldsymbol{\theta}_v\right)\Biggr)\Biggr],
    \label{eq:PPOobjective}
\end{split}
\end{equation}

by tuning $\boldsymbol{\theta}_{p}$ via gradient ascend with Adam \cite{proc:Kingma15}, where: $\epsilon$ is an hyperparameter usually set to $0.2$; $\hat{\mathbb{E}}_e$ is the empirical mean of the advantage function over $N_e$ trajectories; $d\left(s,a,\boldsymbol{\theta}_v\right)$ is the advantage function based on the current critic; $\pi_{\text{old}}\left(a|s,\boldsymbol{\theta}_{p_{\text{old}}}\right)$ is the policy run to collect the $N_e$ trajectories. ``$\text{min}$'' requires to take the minimum (pessimistic bound) between a clipped and an unclipped objective. The unclipped objective is the one that, under a certain constrain, is minimised by the trust region policy optimisation algorithm \cite{proc:Schulman15} that represents the basis of PPO.

Importance sampling is employed due to the difference between the updated policy and the one used to generate $N_e$ trajectories. The clipping operation ``$\text{clip}$'' removes the incentive for changing rapidly the policy making the ratio between $\pi\left(a|s,\boldsymbol{\theta}_{p}\right)$ and $\pi_{\text{old}}\left(a|s,\boldsymbol{\theta}_{p_{\text{old}}}\right)$ moving outside the $\left[1-\epsilon,1+\epsilon\right]$ interval. In the following, this ratio will be indicated by $y\left(\boldsymbol{\theta}_{p}\right)$. The clipping operation allows the definition of the following piecewise probability distribution

\begin{equation}
    \begin{cases}
    \multirow{2}{*}{$\textit{y}\left(\boldsymbol{\theta}_{p}\right) d\left(s,a,\boldsymbol{\theta}_v\right)$} &
    \!\!\!\!\text{for } d\left(s,a,\boldsymbol{\theta}_v\right)\!>\!0 \text{ and } y\left(\boldsymbol{\theta}_{p}\right)\! < \!1\!+\!\epsilon, \\
    & \!\!\!\!\text{or } d\left(s,a,\boldsymbol{\theta}_v\right)\!<\!0 \text{ and } y\left(\boldsymbol{\theta}_{p}\right)\! > \!1\!-\!\epsilon, \\
    & \\
    \left(1+\epsilon\right) d\left(s,a,\boldsymbol{\theta}_v\right) & \!\!\!\! \text{for } d\left(s,a,\boldsymbol{\theta}_v\right)\!>0\! \text{ and } y\left(\boldsymbol{\theta}_{p}\right) \!> \!1\!+\!\epsilon, \\
    & \\
    \left(1-\epsilon\right) d\left(s,a,\boldsymbol{\theta}_v\right) & \!\!\!\! \text{for } d\left(s,a,\boldsymbol{\theta}_v\right)\!<\!0 \text{ and } y\left(\boldsymbol{\theta}_{p}\right) \!< \!1-\!\epsilon ,
    \end{cases}
\end{equation}
whose empirical mean is required to compute $\mathcal{L}_p\left(\boldsymbol{\theta}_p\right)$.

The need of cyclically update $d\left(s,a,\boldsymbol{\theta}_v\right)$ points out the actor--critic scheme of the PPO. The parameters $\boldsymbol{\theta}_{v}$ modelling the critic are tuned via gradient descend by minimising a loss function defined as

\begin{equation}
    \mathcal{L}_v\left(\boldsymbol{\theta}_{v}\right)=\frac{1}{N_t N_e}\sum_{n_t=1}^{N_t}\sum_{n_e=1}^{N_e}\left(v\left(S^{n_e}_{n_t},\boldsymbol{\theta}_{v}\right)-G^{n_e}_{n_t}\right)^2
    \label{eq:criticObjective}
\end{equation}

The corresponding Algorithm is reported in APPENDIX A.

\section{Results}
\label{sec:results}

\begin{figure*}
\centering
\begin{tikzpicture}
  \node (geom)  {\includegraphics[width=110mm]{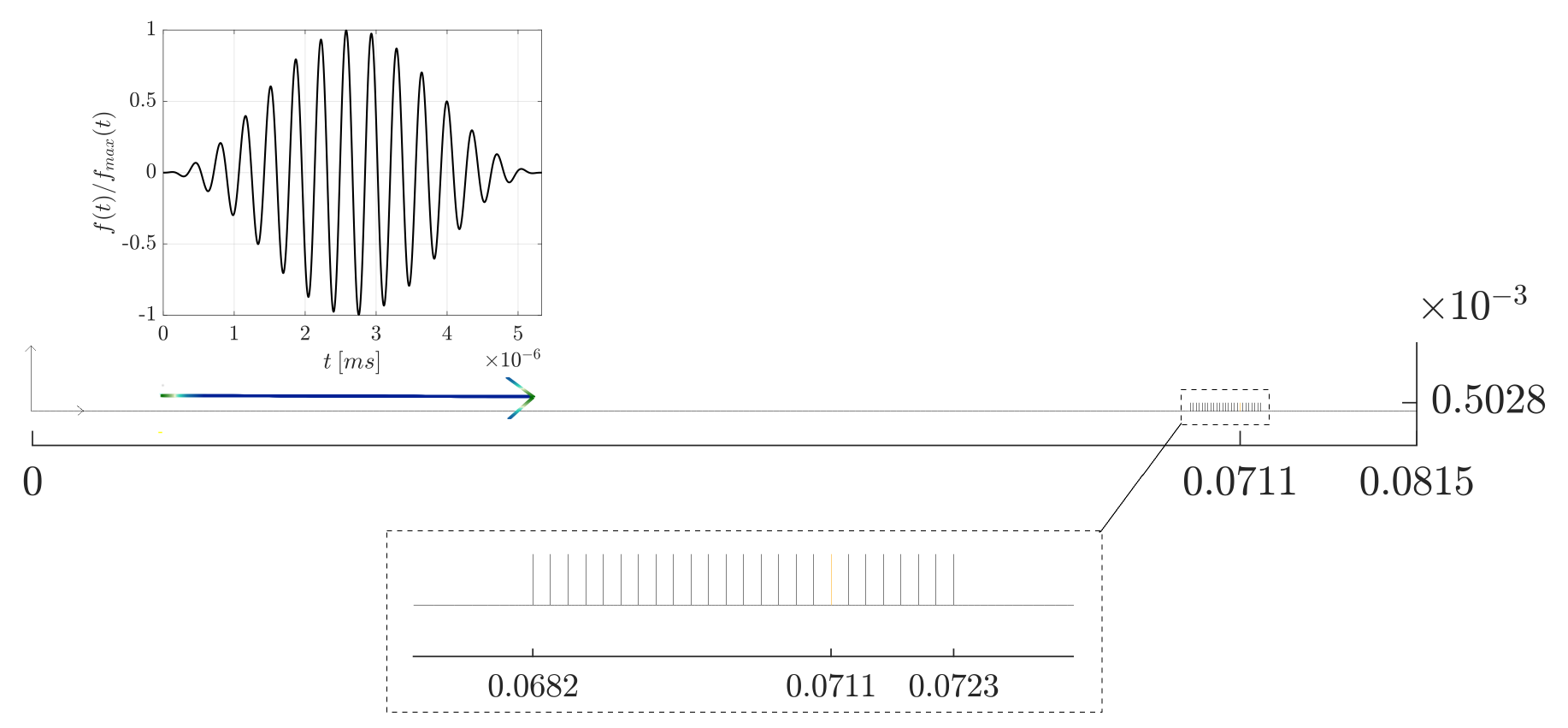}};
  \node[below of=geom,node distance=0cm,xshift=-5.6cm, yshift=0.05cm,font=\color{black}] {\footnotesize $z\text{}\left[\text{m}\right]$};
  \node[below of= geom,node distance=0cm,xshift=-4.9cm, yshift=-0.1cm,font=\color{black}] {\footnotesize $x\text{}\left[\text{m}\right]$};
 \end{tikzpicture}
\caption{\footnotesize Waveguide and resonator array with a magnification of the resonator zone. The harvester has been coloured orange. Top--left corner: applied load, flexural wave packet propagating along the $x$ direction. \label{fig:referenceResConfiguration}}
\end{figure*}

Aim of the work is to show how RL can be applied to metamaterial-based structures for mechanical energy confinement and amplification, highlighting the potentialities of the method as a design optimisation tool in mechanics. Specifically, the aim is to focus and amplify the mechanical energy in a target resonator. Differently from the supervised and the unsupervised paradigms, RL does not require to assemble a data set a priori. Agent--environment interactions are collected on the fly by simulating wave propagation in the system through the Finite Element Method (FEM).

Both the resonators and the waveguide, depicted in Fig. \ref{fig:referenceResConfiguration}, are made by aluminium with density $\rho=2710 \text{ kg}/\text{m}^3$ and Young's modulus $E=70 \text{ GPa}$. The cross sectional area of the waveguide is $B_w=3\cdot 10^{-6} \text{ m}^2$, while the moment of inertia appearing in the dispersion relation of the propagating flexural wave is $I_w=2.5\cdot 10^{-13} \text{ m}^4$. The resonator moment of inertia $I_r$ involved in the first bending mode excited by the wave passage is $I_r=0.4909\cdot 10^{-13} \text{ m}^4$, while the resonator cross sectional area is $B_r=0.785\cdot 10^{-6}$. The contribution of no other bending mode has been considered. The bending frequency featuring the first bending mode is computed as

\begin{equation}
\omega_r=1.875^2\sqrt{\frac{E I_r}{\rho B_r L_r^5}},
\label{eq:firstBendingMode}
\end{equation}
where $L_r$ is the resonator length.

The metastructure is excited through the wave packet depicted in Fig. \ref{fig:referenceResConfiguration}. The frequency content of the wave packet coincides with the angular frequency of the target resonator first bending mode $\omega_h=17.67$ MHz, obtained by substituting the length $L_h=5.028\cdot 10^{-4}$ in Eq. \ref{eq:firstBendingMode} to $L_r$. The same type of excitation was provided through an electrodynamic shaker in an experimental setting by \cite{art:APL20}. The wave number $\kappa_w=3470 \text{ rad/m}$ is determined through the following dispersion relation
\begin{equation}
    \omega = \sqrt{\frac{E I_w}{\rho B_w}}\kappa^2,
    \label{eq:dispersionFlex}
\end{equation}
by setting $\kappa=\kappa_w$ and $\omega_h=\omega$, see \cite{chp:DePonti2021}. Consequently, the wave length $\lambda_w$ is equal to $1.8\cdot10^{-3}$ m.

To avoid wave reflections, two absorbing layers with length $L_w^{\text{abs}}=5\lambda_w=9.1\cdot10^{-3}$ m have been implemented at the extremities of the waveguide as in \cite{art:Rajagopal12}. Comparable absorbing properties were experimentally obtained through acoustic black holes \cite{art:OBoy10,art:Georgiev11,art:APL20}. Due to the presence of the absorbing layers, the exciting force has been applied at $x=9.1\cdot10^{-3}$ m. Different FE discretisations have been employed for the considered optimisation cases. However, every discretisation has employed five FEs in between neighbouring resonators to catch localised effects due to the resonator interaction. Resonators have been modelled using a lumped spring--mass representation. Wave propagation has been simulated for $T=1.25\cdot 10^{-5}$ s for the first two analyzed optimisation setting. The second case, involving the grading of both the resonator lengths and spacing, will require further discussions to set $T$. A Rayleigh damping featuring a ratio $\xi$ roughly of $0.2\%$ for the first $40$ structural modes has been considered. Mechanical data are collected in Tab. \ref{tab:optimisationDataMech}.

\begin{table}[h]
\tiny
    \centering
    \[
    \begin{array}{lrlr}
    \toprule
    & \multicolumn{2}{c}{\mbox{Material: aluminium}} & \\
    \midrule
    \multicolumn{1}{l}{\mbox{Density}} &  & \multicolumn{1}{l}{\rho=2710 \text{ kg}/\text{m}^3} & \\
    \multicolumn{1}{l}{\mbox{Young's modulus}} & & \multicolumn{1}{l}{E=70 \text{ GPa}} & \\
    \multicolumn{1}{l}{\mbox{Rayleigh damping ratio}} & & \multicolumn{1}{l}{\approx\xi=0.2\%} & \multicolumn{1}{l}{\mbox{ on the first }40\mbox{ modes.}}\\
    \midrule
    \multicolumn{2}{c|}{\mbox{Waveguide geometry}} & \multicolumn{2}{c}{\mbox{Resonator geometry}} \\
    \midrule
    \mbox{Moment of inertia} & \multicolumn{1}{r|}{I_w=2.5\cdot 10^{-13} \text{ m}^4} & \mbox{Moment of inertia} & I_r=0.4909\cdot 10^{-13} \text{ m}^4\\
    \mbox{Cross sectional area} & \multicolumn{1}{r|}{B_w=3\cdot 10^{-6} \text{ m}^2} & \mbox{Cross sectional area} & B_r=0.785\cdot 10^{-6} \text{ m}\\
    \mbox{Total Length} & \multicolumn{1}{r|}{L_w=0.815 \text{ m}} & \mbox{Target resonator length} & L_h=5.028\cdot 10^{-4} \text{ m} \\
    \mbox{Absorbing layer length} & \multicolumn{1}{r|}{L_w^{\text{abs}}=9.1\cdot10^{-3}\text{ m}} & & \\
    \bottomrule
    \end{array}
    \]
    \caption{Metasurface mechanical data.}
    \label{tab:optimisationDataMech}
\end{table}

\normalsize The input force generates two wave fronts, one propagating left (towards points with smaller $x$), and the other propagating right. The left propagating wave is immediately damped out by the absorbing layer, while the other wave propagates until it reaches the second absorbing layer. The wave travels with group velocity $c_g$ equal to $1.018\cdot 10^4$ m$/$s determined by substituting Eq. \ref{eq:dispersionFlex} in $c_g=\partial \omega / \partial \kappa$.

In Sec. \ref{sec:dispersionRelation}, the dispersion relation for a local resonant waveguide is analytically derived to get a mechanical insight on the optimisation problem. Outcomes concerning the optimisation procedure are reported: in Sec. \ref{sec:resultsLengths} with respect to the grading of the resonator lengths; in Sec. \ref{sec:resultsLengthSpacing}, with respect to the resonator lengths and spacing. For sake of clarity, interpolation points will be differently labelled: with $z$--IP (``IP'' refers to interpolation point), when they refer to the resonator lengths; with $\zeta$--IP, when they set the resonator spacing.

\subsection{Dispersion relation for 1D rainbow-based metamaterials}
\label{sec:dispersionRelation}

A dispersion relation expresses the dependence between $\omega$ and $\kappa$ for a wave propagating in a certain periodic medium. Provided the grading is gentle enough, the global behaviour of a rainbow metamaterial is deduced from the local dispersion curves of the constituent elements \cite{art:Colombi2016}. This allows to compute the dispersion relation for a transverse wave in the rainbow metamaterial by formulating the infinitesimal equilibrium equation of the periodic problem, according to the Euler-Bernoulli beam theory

\begin{equation}
    E I_w \frac{\partial^4 u_w(x,t)}{\partial x^4} + \rho B_w \frac{\partial^2 u_w(x,t)}{\partial t^2} = 0,
    \label{eq:EBtransverseEq}
\end{equation}
where $u_w(x,t)$ is the waveguide displacement field (being $E$, $I_{\omega}$, $\rho$, $B_{\omega}$ the Young modulus, moment of inertia, mass density and cross section of the waveguide, respectively), and by employing a lumped mass--stiffness model to include the presence of resonators.

To avoid boundary effects, the length of the waveguide has been assumed to be infinite. By assuming that the motion of the resonators is dependent only on the motion of the underlying waveguide, it is possible to write:

\begin{equation}
 E I_w \frac{\partial^4 u_w\left(x,t\right)}{\partial x^4} + \rho B_w \frac{\partial^2 u_w\left(x,t\right)}{\partial t^2} =
 \sum_{i=-\infty}^{\infty} K_r^i \left(u_r^i\left(t\right) - u_w\left(x,t\right)\right) \delta \left(x-iD\right) ,
\label{eq:localResEBequilibrium}
\end{equation}
where: $u_r^i\left(t\right)$ is the displacement of the $i$th mass of the local resonance system; $K_r^i$ is the lumped stiffness of the spring that connects the lumped mass $M_r^i$ with the underlying waveguide; $D$ is the distance between two resonators; $\delta\left(x-iD\right)$ is the Dirac delta.

To solve the system, the solution of the equation of motion of the $i$th resonator is needed. This is formulated as follows:
\begin{equation}
M^i_r \frac{\partial^2 u^i_r\left(t\right)}{\partial t^2} + K^i_r \left(u^i_r\left(t\right) - u_w\left(iD,t\right)\right) = 0,
\label{eq:EqMotionRes}
\end{equation}
granted that the stiffness $K_r=K_r^i$, the mass $M_r=M_r^i$, and the distance $D$ are assumed the same, thanks to the locally periodic behaviour of the graded metamaterial.

In this work, $M_r=\rho B_r L_r$, while $K_r=\omega_r^2 M_r$, where $\omega_r$ has been computed as in Eq. \eqref{eq:firstBendingMode}. In case of identical resonators, the solution obtained by \cite{art:Skelton18} is:
\begin{equation}
    \left(\sum_{i=-\infty}^{\infty} \frac{1}{EI_w (\kappa - ig)^4 - \rho B_w \omega^2}\right)^{-1}- \frac{K_r\omega^2}{D\left(\frac{K_r}{M_r}-\omega^2\right)} = 0,
    \label{eq:Skelton}
\end{equation}
where $g = \frac{2\pi}{D}$ is the reciprocal lattice vector, $\kappa$ the wavenumber and $\pi$ is just the number.

The summation in Eq. \eqref{eq:Skelton} is needed to model the Bragg effect that develops near to the boundary of the First Brillouin Zone (FBZ). If the local resonance band gap is sufficiently separated from the Bragg one, i.e. they arise at different frequencies, this effect can be neglected as it marginally contributes in defining the shape of the dispersion relation, leading to simplify Eq. \eqref{eq:Skelton} as 
\begin{equation}
    EI_w \kappa^4 - \rho B_w \omega^2 - \frac{K_r\omega^2}{D\left(\frac{K_r}{M_r}-\omega^2\right)} = 0.
    \label{eq:simplifiedSkelton}
\end{equation}

In turn, this equation can be reshaped to show the dependence of $\kappa$ over the lattice spacing, obtaining:
\begin{equation}
   \kappa =\mathcal{W}(\omega) + \frac{1}{\sqrt[4]{D}} \mathcal{R}(\omega,K_r,M_r),
   \label{eq:DispersionRelation}
\end{equation}
for a fixed frequency $\omega$, stiffness $K_r$ and mass $M_r$ of the resonators. It is interesting to see that $\mathcal{W}$ contains all the contributions that define the dispersion relation for the infinite beam, while $\mathcal{R}$ contains the contribution brought by the resonators. The distance $D$ is kept out of $\mathcal{R}$ to show how it affects $\kappa$.

Eq. \eqref{eq:DispersionRelation} highlights which are the design parameters to be optimised for a proper grading: the resonator stiffness $K_r$, appearing in the formulation of $\mathcal{R}$; the distance $D$ in between the resonators. From a geometric point of view, setting $K_r$ is allowed by setting the resonator length $L_r$. 
In light of that: Sec. \ref{sec:resultsLengths} will investigate how to optimally grade the resonator lengths $L_r$  or spacing $D$; Sec. \ref{sec:resultsLengthSpacing} how to optimally combined both $L_r$ and $D$

\subsection{Optimisation of the resonator lengths or spacing}
\label{sec:resultsLengths}

The optimisation procedure detailed in Sec. \ref{sec:methodology} has been applied to set the lengths, and therefore the lumped mass and stiffness properties, of $25$ resonators. Resonators have been spaced by roughly $\lambda_w/11$, within the range  $\left[\lambda/15,\lambda/2\right]$ for which resonator interaction is expected at the subwavelength scale \cite{art:Lemoult11}. The precise value depends on the employed discretisation. Specifically, the waveguide has been discretised through $376$ Euler Bernoulli FEs \cite{book:Belytschko00} with length equal to $0.0344\cdot10^{-3}$ m where the resonators are placed, and equal to $0.344\cdot10^{-3}$ m outside that zone. The integration time step $\Delta t$ has been set to $3\cdot10^{-9}$ s. The 
target resonator coincides with the $18$th resonator from the left. Its height has been fixed to $L_h$.

\begin{figure}[h!]
\centering
\includegraphics[width=0.5\columnwidth]{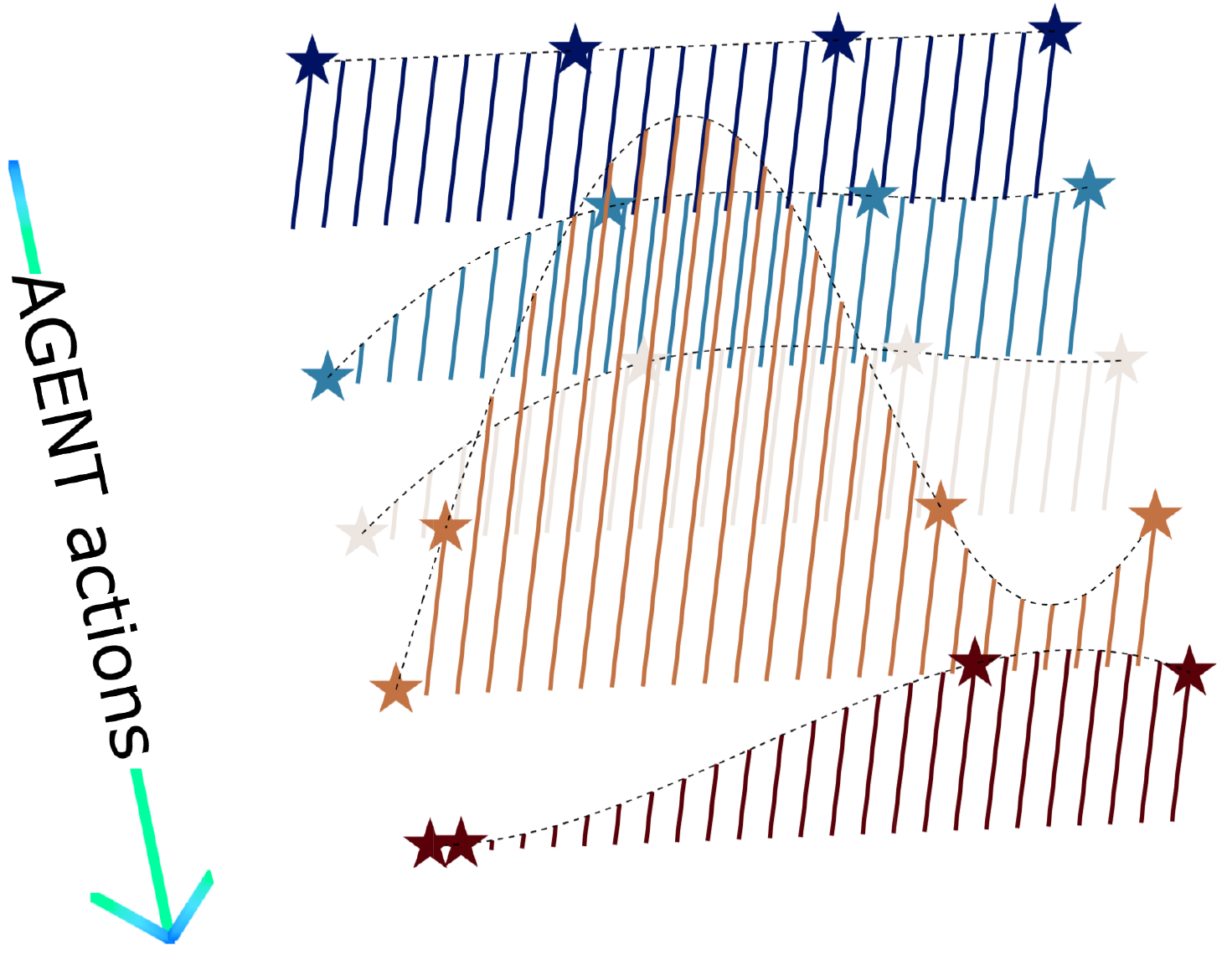}
\caption{\footnotesize Optimisation of the resonator lengths: sequence of actions leading to the configuration with highest reward.\label{fig:sequenceResonatorHeight}}
\end{figure}

Four points have been used to define the envelope $z=p\left(x\right)$ of the resonator lengths. In Fig. \ref{fig:interpolationExample}, the four points have been depicted by star markers. Specifically, the $3$rd $z$--IP has always been placed at the tip of the target resonator, while the $1$st and $4$th points have been forced to match the tips of the first and of the last resonators, respectively. Therefore, $N_s=4$ continuous variables have been exploited, coinciding in the $x$--$z$ plane with the ordinates of the $1$st and $4$th $z$--IPs, and with the $\left(x,z\right)$ coordinates of the $2$nd $z$--IP. The agent has been asked for setting these continuous variables by taking the following sequence of $N_t=4$ actions: (i) set the $z$ coordinate of the $1$st $z$--IP; (ii) set the $z$ coordinate of the $4$th $z$--IP; (iii) set the abscissa and then the ordinate (iv) of the $2$nd $z$--IP. The order of operations is exemplified in Fig. \ref{fig:sequenceResonatorHeight} and schematically reported in Tab. \ref{tab:agentActionsLengths}.

The configuration space has been constrained on the basis of the physical understanding of the system acquired in \cite{art:NJP20,art:APL20}. No other a priori knowledge has been exploited. For example, no notion has been employed to set the system configuration at the start of each trajectory. Indeed, all resonators have been set equal to $L_h$ at each trajectory start, as depicted in the top--left plot in Fig. \ref{fig:sequenceResonatorHeight}.

The $z$--IP ordinates have been allowed for varying in the $\left[0,L^{\text{max}}_r\right]$ interval, where $L^{\text{max}}_h=9.156\cdot 10^{-4} \text{ m}$. The rationale behind this value is that if a sinusoidal force with frequency equal to $\omega^{\text{max}}_r$ is used to excite a single degree of freedom system centered at $\omega_h$, a $10\%$ attenuation will be produced, with $\omega^{\text{max}}_r$ computed as in Eq. \eqref{eq:firstBendingMode} by setting $L_r=L^{\text{max}}_r$. The $x$ coordinate of the second $z$--IP have been allowed to move between the first resonator position and the target resonator position. A description of the agent actions is reported in Tab. \ref{tab:agentActionsLengths}. Cubic B--splines have been exploited to define $p\left(x\right)$. If a resonator length results smaller than $L_h/20$, this resonator is removed from the design. This threshold value has been set to avoid the ill conditioning of the stiffness matrix caused by very high stiffness of the lumped resonators. In this way, the procedure can possibly reduce the number of resonators.

\begin{table}[h!]
\footnotesize
    \centering
    \[
    \begin{array}{cccc}
    \toprule
    \mbox{Action} & \mbox{What} & \mbox{Variable value} & \mbox{Range of} \\
    \mbox{ordering} & \mbox{is modified} & \mbox{starting state} & \mbox{possible values}\\
    \midrule
    1 & 1\mbox{st $z$--IP, }z & 5.028\cdot 10^{-4} & \left[0,9.156\cdot 10^{-4} \right]\text{ m} \\
    2 & 4\mbox{th $z$--IP, }z & 5.028\cdot 10^{-4} & \left[0,9.156\cdot 10^{-4}\right]\text{ m} \\
    3 & 2\mbox{nd $z$--IP, }x & 0.0697 \mbox{ m} & \left[0.0682,0.0711\right]\mbox{ m} \\
    4 & 2\mbox{nd $z$--IP, }z & 5.028\cdot 10^{-4} & \left[0,9.156\cdot 10^{-4}\right]\text{ m} \\
    \bottomrule
    \end{array}
    \]
    \caption{Optimisation of the resonator lengths: description of the agent actions. The interpolation points are the ones reported with cyan markers in Fig. \ref{fig:interpolationExample}.}
    \label{tab:agentActionsLengths}
\end{table}

\normalsize A NN featuring two fully connected layers with just $32$ neurons and $N_v=1348$ tunable weights has been employed to approximate the advantage function
\begin{equation}
d\left(s,a,\boldsymbol{\theta}_v\right)=\mathcal{F}_{32-4}\circ\mathcal{F}_{4-32}\left(s,a,\boldsymbol{\theta}_v\right),
    \label{eq:advFunctionNN}
\end{equation}
where $\mathcal{F}_{N_n-N_m}:\mathbb{R}^{N_n}\rightarrow\mathbb{R}^{N_m}$ is the transformation induced by a fully connected layer receiving a vector of length $N_n$ in input and returning a vector with length $N_m$ in output. 
A similar NN architecture, exploiting $N_p=1282$ parameters, has been used to calibrate the policy parameters
\begin{equation}
\left[\mu\left(s,\boldsymbol{\theta}_p\right),\sigma\left(s,\boldsymbol{\theta}_p\right)\right]=\mathcal{F}_{32-2}\circ\mathcal{F}_{4-32}\left(s,a,\boldsymbol{\theta}_p\right).
\label{eq:policyParamNN}
\end{equation}

\begin{figure}
\footnotesize
\centering
\begin{tikzpicture}
\node (geom)  {\includegraphics[width=70mm]{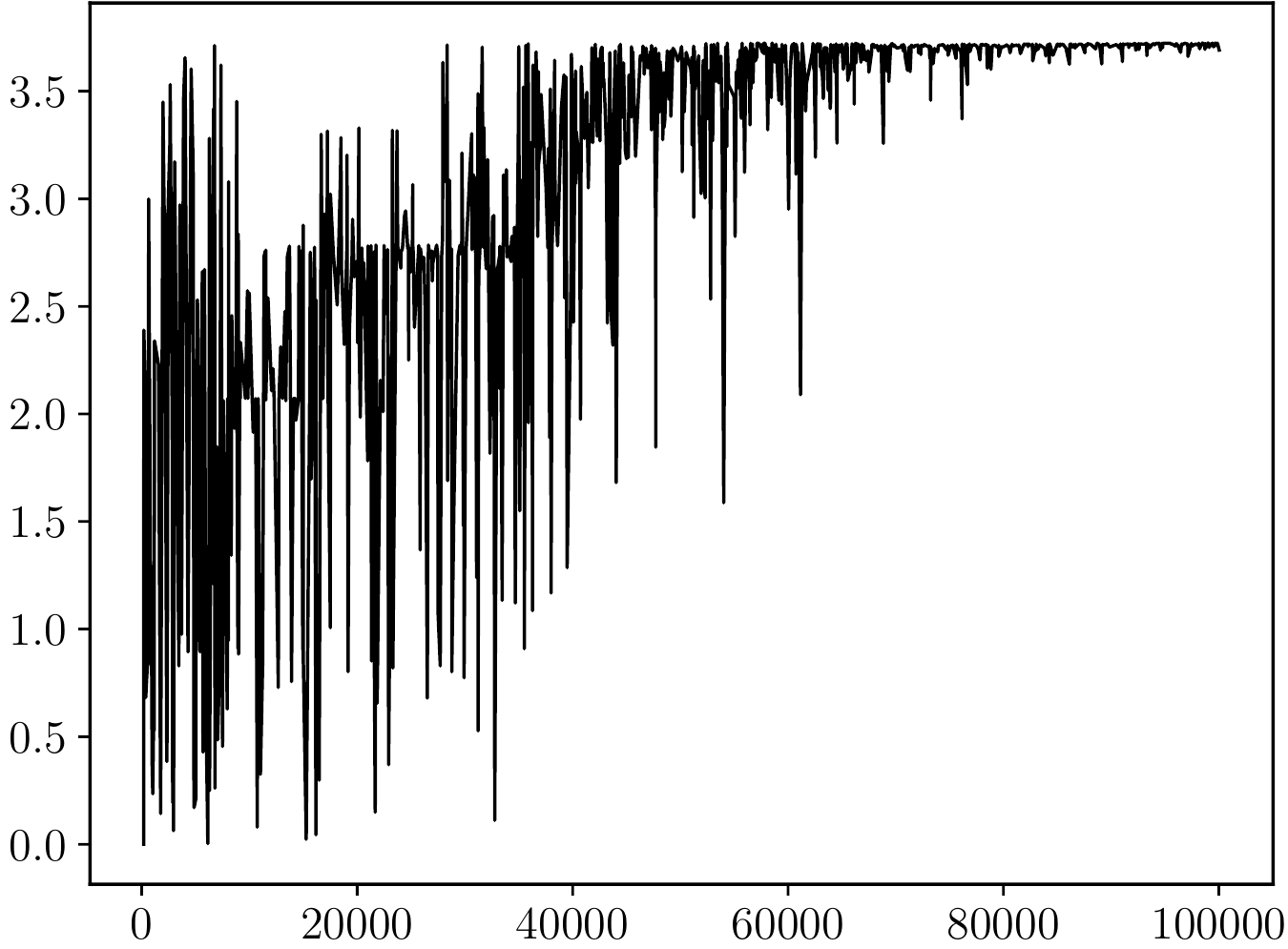}};
\node[below of=geom,node distance=0cm,xshift=-4cm, yshift=0cm,rotate=90,font=\color{black}] {$R_{N_t}/R^H_{N_t}$};
\node[below of=geom,node distance=0cm,xshift=0cm, yshift=-3cm,font=\color{black}] {$n_i$};
\end{tikzpicture}
\caption{Optimisation of the resonator lengths: evolution of the final episode reward during the agent training divided by $R^H_{N_t}$.}
\label{fig:6_1episodeReward}
\end{figure}

Performance has not been improved either by adding hidden layers or by increasing the number of neurons. A number $N_e=32$ of trajectoris have been collected for each estimate of the PPO objective $\mathcal{L}_p$ expressed in Eq. \eqref{eq:PPOobjective}. It has been verified that a smaller $N_e$ deteriorates the performance of the procedure, and that a larger $N_e$ increases the computational effort without improving the procedure outcome. The total number of agent--environment interactions has been fixed to $N_{i}=100,000$ FE simulations. As it will be assessed by Fig. \ref{fig:6_1episodeReward}, this number has guaranteed the convergence to a quasi--deterministic policy obtained by greatly reducing $\sigma\left(s,\boldsymbol{\theta}_p\right)$ with respect to $\mu\left(s,\boldsymbol{\theta}_p\right)$ for some state $s$. The learning rate $\eta$ has been set to $5\cdot10^{-4}$ after a trial and error procedure. The employed function approximators and hyperparameters are collected in Tab. \ref{tab:optimisationData}. The stable baselines library \cite{code:stableBaselines} has been exploited for the implementation of the PPO algorithm.

\begin{table*}[h!]
\footnotesize
    \centering
    \[
    \begin{array}{lc}
    \toprule
    \multicolumn{2}{c}{\mbox{Function approximators}} \\
    \midrule
    \mbox{Advantage function approximator} & d\left(s,a,\boldsymbol{\theta}_v\right)=\mathcal{F}_{32-4}\circ\mathcal{F}_{4-32}\left(s,a,\boldsymbol{\theta}_v\right) \\
    \mbox{Function calibrating the policy parameters} & \left[\mu\left(s,\boldsymbol{\theta}_p\right),\sigma\left(s,\boldsymbol{\theta}_p\right)\right]=\mathcal{F}_{32-2}\circ\mathcal{F}_{4-32}\left(s,\boldsymbol{\theta}_p\right) \\
    \midrule
    \multicolumn{2}{c}{\mbox{Hyperparameters}} \\
    \midrule
    \mbox{Number of trajectories for the }\mathcal{L}_p \mbox{ estimation} & N_e=32\\
    \mbox{Number of agent--environment interactions} & N_i=100,000 \\
    \mbox{Learning rate} & \eta=5\cdot 10^{-4}\\
    \bottomrule
    \end{array}
    \]
    \caption{Optimisation of the resonator lengths: PPO algorithm approximating functions and hyperparameters.}
    \label{tab:optimisationData}
\end{table*}

\normalsize In Fig. \ref{fig:sequenceResonatorHeight}, the sequence of actions leading to the configuration with the highest reward is reported. As it can be observed, the $2$nd $z$--IP has been moved very close to the $1$st $z$--IP, so much that the resulting curve could have been obtained by interpolating $3$ points only. This further confirms that the adopted state description has reduced the space of possible designs but including the most effective configurations for mechanical energy confinement and amplification.

The convergence to a deterministic policy is assessed in Fig. \ref{fig:6_1episodeReward}, where the evolution of the episode final reward $R_{N_t}$ during the agent training is depicted. The graph ordinate is normalised with respect to the reward $R^H_{N_t}$ collected for the waveguide equipped with the target resonator only, and it is plotted against the number $n_{i}=1,\ldots,N_{i}$ of agent--environment interactions. The ratio $R_{N_t}/R^H_{N_t}$ quantifies the advantage of adopting a certain metastructure design with respect to the configuration featuring a single resonator.

\begin{figure}
    \centering
    \subfloat[Single resonator configuration, $R_{N_t}/R^{H}_{N_t}\!=\!1.000$.\label{fig:onlyHarvester25}]{\includegraphics[width=40mm]{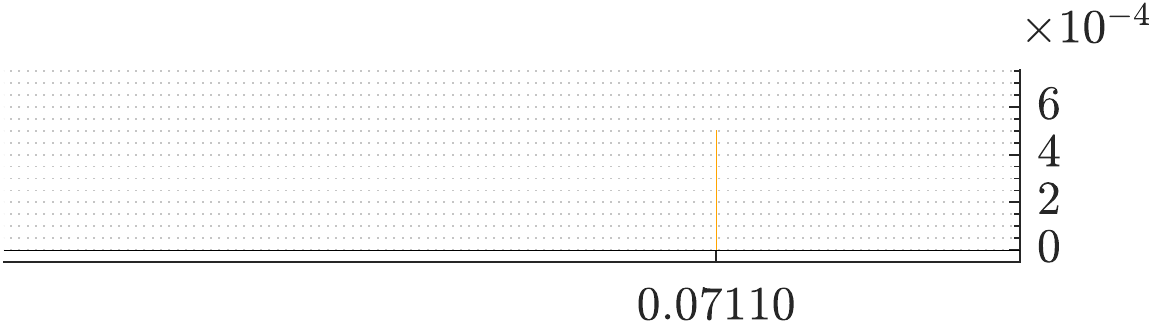}}\hspace{10mm}
    \subfloat[Reference configuration (linear grading of the resonator lengths), $R_{N_t}/R^{H}_{N_t}=3.504$.\label{fig:linearGrading25}]{\includegraphics[width=40mm]{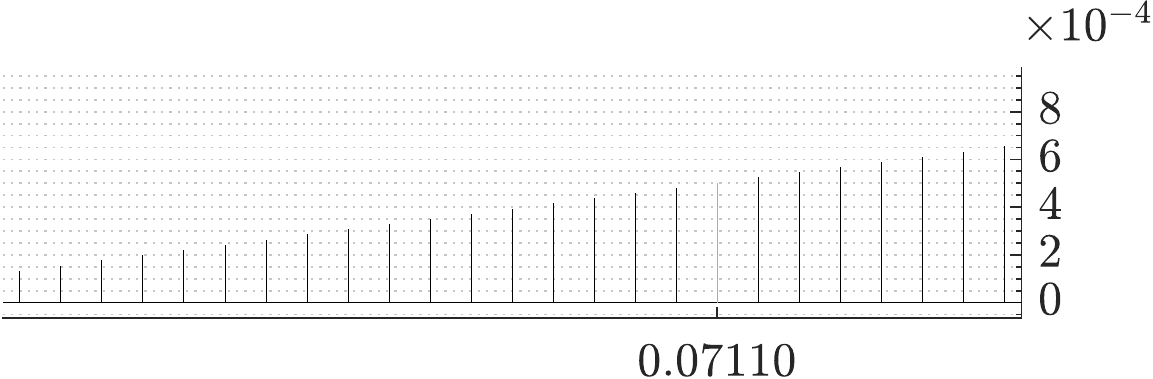}} \\
    \subfloat[Best RL discovered configuration, $R_{N_t}/R^{H}_{N_t}=3.669$.\label{fig:bestConfigurationLengths}]{\includegraphics[width=40mm]{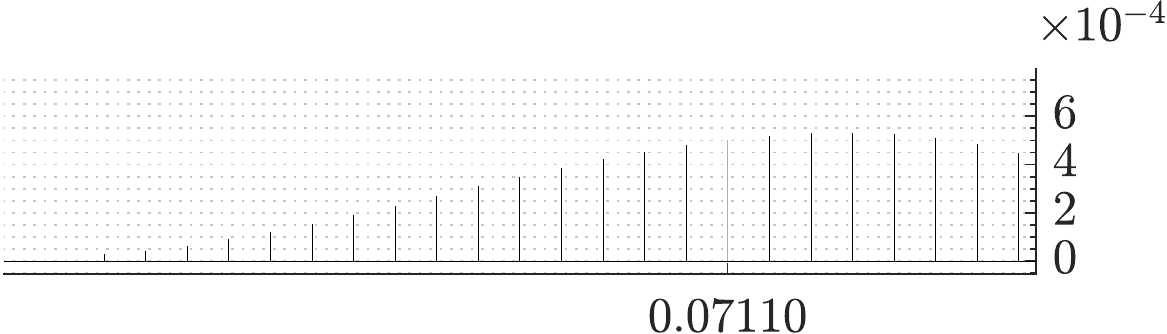}}\hspace{10mm}
    \subfloat[Converged RL policy configuration, $R_{N_t}/R^{H}_{N_t}=3.537$.\label{fig:testConfigurationLengths}]{\includegraphics[width=40mm]{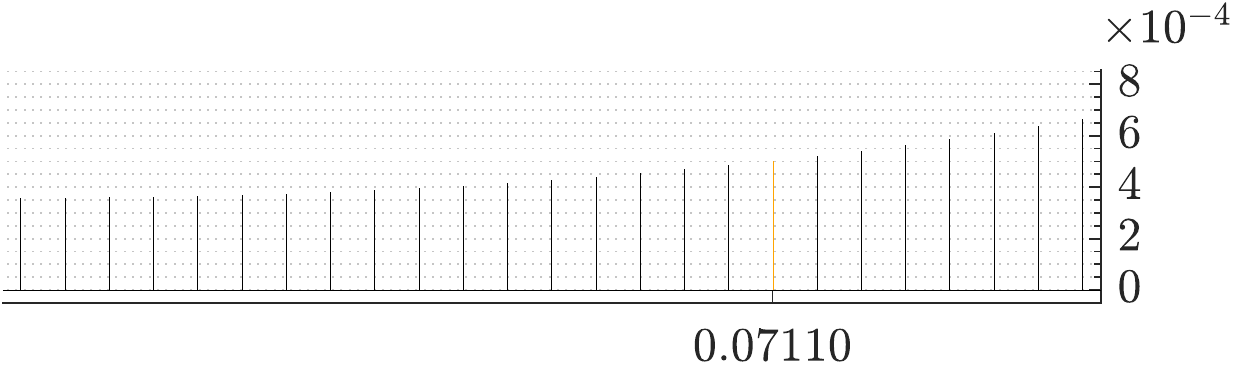}}
    \caption{Optimisation of the resonator lengths: resonator arrangements for the reference and for the optimised configurations.}
\end{figure}

\normalsize The deterministic policy obtained after $N_i$ agent--environment interactions generates a configuration, depicted in Fig. \ref{fig:testConfigurationLengths}, featuring a $R_{N_t}/R^H_{N_t}$ ratio equal to $3.537$. This ratio is slightly smaller (by $\approx 3.5\%$) than the $R_{N_t}/R^H_{N_t}=3.669$ ratio associated to the configuration obtained through the action sequence depicted in Fig. \ref{fig:sequenceResonatorHeight}. In other words, the best generated configuration has been produced before the total number of policy updates have been performed. The performance of the optimisation procedure is evaluated not just by comparing $R_{N_t}$ with $R^H_{N_t}$, but also $R_{N_t}$ with the reward featuring the linear grading rule shown in Fig. \ref{fig:linearGrading25}, as proposed in \cite{art:NJP20}.

As it can be seen, the best RL discovered configuration largely enhances the 
mechanical energy confinement and amplification capability with respect to $R^H_{N_t}$. This enhancement has been obtained constraining the design space as unique a priori knowledge guiding the agent actions. The RL agent has been able not only to recover what previously known, but also to overcome the performance of the linear grading resonator arrangement by $4.71\%$. Despite the limited increase, no further major improvements are expected by playing only on tuning the resonator lengths, as also shown by the work of \cite{ErturkGrading2022} when comparing different linear and parabolic grading rules.

The lower performance of the finally obtained deterministic policy with respect to the best discovered configuration does not mean that the MDP solution has been unsuccessful or meaningless. On the contrary, the utility of the proposed approach has been confirmed: by the convergence of the PPO algorithm to a suboptimal policy generating a resonator arrangement close in terms of $R_{N_t}/R^H_{N_t}$ to the best configuration found during the algorithm explorations; by the fact that the best discovered configuration has been produced after the reduction of the space of possible configurations due to the first policy updates. Moreover, arriving to a suboptimal policy is the usual outcome of a RL procedure unless of a dramatic increasing in computing time \cite{book:Sutton18}.

Some theoretical arguments discussing how to theoretically find an optimal grading are reported to further assess the outcome of the optimisation procedure. A possible way to approach the problem is to think that the propagating energy of the wave is progressively reflected by the resonators. The matter becomes to define the intensity of such reflection and to find a grading law that minimises them. With this goal, we consider the waveguide as made of a series of layers of different homogeneous material whose properties are given by the dispersion relation stated in Eq. \ref{eq:Skelton}. Therefore, it is possible to define a reflection coefficient $\varrho_i$ between each layer, so between each resonator cell and the preceding one, in the following way:
\begin{equation}
    \varrho_{i} = \frac{k_i -k_{i+1}}{k_{i+1} +k_i}
\end{equation}
and a transmission coefficient $\tau_i$ that is:
\begin{equation}
    \tau_{i} = \frac{2k_i}{k_{i+1} +k_i},
\end{equation}
Where $k_i$ is the wavevector associated to the $i$th layer, and a resonator cell is defined by a resonator and by the attached waveguide portion.

These equations are the same used in optics, and they stem from imposing both displacements and force continuity at the interfaces between the layers. Physically, we are interested in the square of the transmission coefficients from one layer to the next, namely transmittance and reflectance. In these terms, the optimisation problem can be thought as finding which sequence of layers maximises the wave intensity on the desired final layer of the target resonator, i.e the one that maximises the product of all the layer transmittance. The best RL configuration (Fig. \ref{fig:bestConfigurationLengths}) partially fulfills such condition, with an average variation of $\Delta \kappa \simeq 174\:\text{rad}/\text{m}$ for the initial portion of the array. On the other hand, this approach fails close to the target resonator, since it is not able to predict the real wavefield generated by the multiple scattering and emissions of the resonators. In particular, as explained in the last section, the effectiveness of rainbow metastructures is strongly related to the interaction time between the wave and the target resonator. This means that, the maximum peak power provided by a proper tuning of the transmittance, may not be the optimal solution to harvest energy in a given period of time. 

The deterministic context in which the optimisation problem has been set allows for considering the best configuration discovered during the exploration phase as the outcome of the procedure. As the number of agent--environment interactions $n_i$ run up that point has been quite limited ($\approx 5000$), the method seems suitable to investigate mechanical problems involving more time demanding FE simulations. Relying on the configuration related to the policy finally determined by the PPO algorithm would have been necessary if any source of stochasticity, e.g. related to the loading conditions, had been included in the FE simulations. In that case, the best configuration could have not be generated by the sequence of actions leading to the highest $R_{N_t}$ during the policy exploration phase, as that value of $R_{N_t}$ may have been connected to special loading conditions.


The interest is now directed to optimise the spacing in between the resonators. Eq. \eqref{eq:DispersionRelation} directly shows the dependence of the wavevector with respect to just a change in distance in the lateral resonators. This is also true considering that an enlargement of the cell is reducing the influence of the resonators over the dispersion relation, while smaller spacing between resonators straightens the contribution of $\mathcal{R}$.

A RL agent has been asked for optimally ruling the resonator spacing for mechanical energy confinement and amplification. The minimum spacing has been set to $\lambda/16$, the maximum spacing to $\lambda/2$. A discrete number of spacing levels ($\lambda/16,\lambda/8,3\lambda/16\ldots$) has been allowed by increasing the minimum distance by $\lambda/16$. A number $N_s=6$ of continuous variables setting the coordinates of four $\zeta$--IPs in the $\left(x,\zeta\right)$ plane and ruling the envelope $\zeta=p\left(x\right)$ has been employed to constrain the design space. The $\zeta$ coordinates have been allowed to vary continuously, operating the discretisation of the spacing levels in a second time. 
Tab. \ref{tab:agentActionsSpacing} resumes the continuous variables possible range of variations and starting values clarifying the ordering of the agent actions. Concerning the resonator lengths, a linear grading rule has been assumed. The same tuning parameters employed for the PPO algorithm in Sec. \ref{sec:resultsLengths} have been exploited, still modelling $d\left(s,a,\boldsymbol{\theta}_v\right)$ and calibrating the policy parameters $\left[\mu\left(s,\boldsymbol{\theta}_p\right),\sigma\left(s,\boldsymbol{\theta}_p\right)\right]$ through two different NNs featuring two hidden layers with $32$ neurons.

\begin{table}[h!]
\footnotesize
    \centering
    \[
    \begin{array}{cccc}
    \toprule
    \mbox{Action} & \mbox{What} & \mbox{Variable value} & \mbox{Range of} \\
    \mbox{ordering} & \mbox{is modified} & \mbox{starting state} & \mbox{possible values}\\
    \midrule
   1 & 1\mbox{st $\zeta$--IP, }\zeta & \lambda/16 & \left[0,\lambda/2\right] \\
    2 & 4\mbox{th $\zeta$--IP, }\zeta & \lambda/16 & \left[0,\lambda/2\right] \\
    3 & 2\mbox{nd $\zeta$--IP, }x & 0.0696 \mbox{ m} & \left[0.0683,0.0703\right]\mbox{ m} \\
    4 & 2\mbox{nd $\zeta$--IP, }\zeta & \lambda/16 & \left[0,\lambda/2\right] \\
   5 & 3\mbox{rd $\zeta$--IP, }x & 0.0710 & \left[0.0703,0.0723\right]\mbox{ m} \\
    6 & 3\mbox{rd $\zeta$--IP, }\zeta & \lambda/16 & \left[0,\lambda/2\right] \\
   \bottomrule
    \end{array}
    \]
    \caption{Optimisation of the resonator spacing: description of the agent actions. The $\zeta$--IPs are reported with cyan markers in Fig. \ref{fig:8_3geom}.}
    \label{tab:agentActionsSpacing}
\end{table}

\begin{figure}
    \centering
    \subfloat[Best RL discovered configuration, $R_T/R^{H}_T=3.600$ .\label{fig:8_3geom}]{\includegraphics[width=55mm]{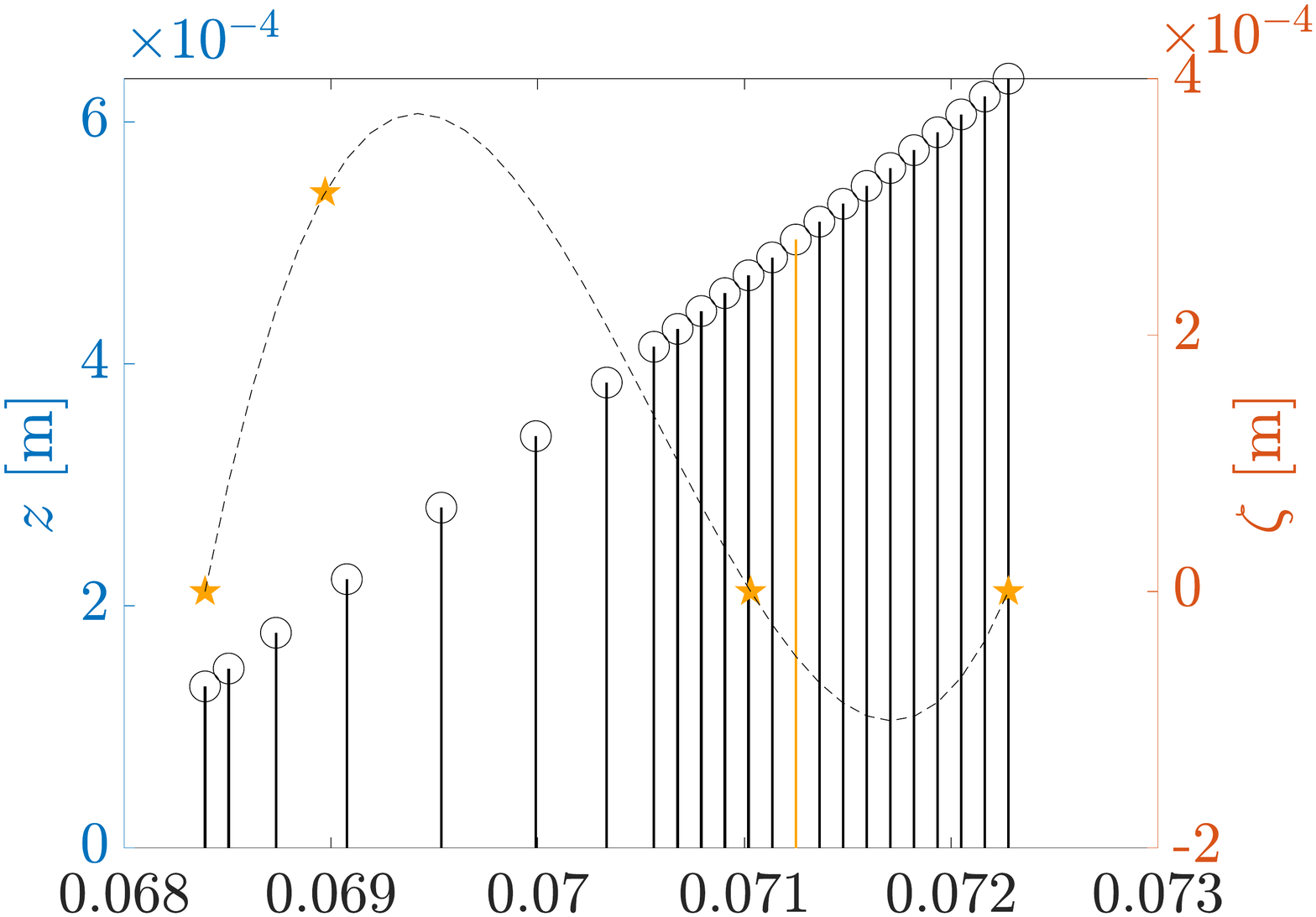}}
    \subfloat[Spectrogram.\label{fig:8_3Spectrogram}]{\includegraphics[width=55mm]{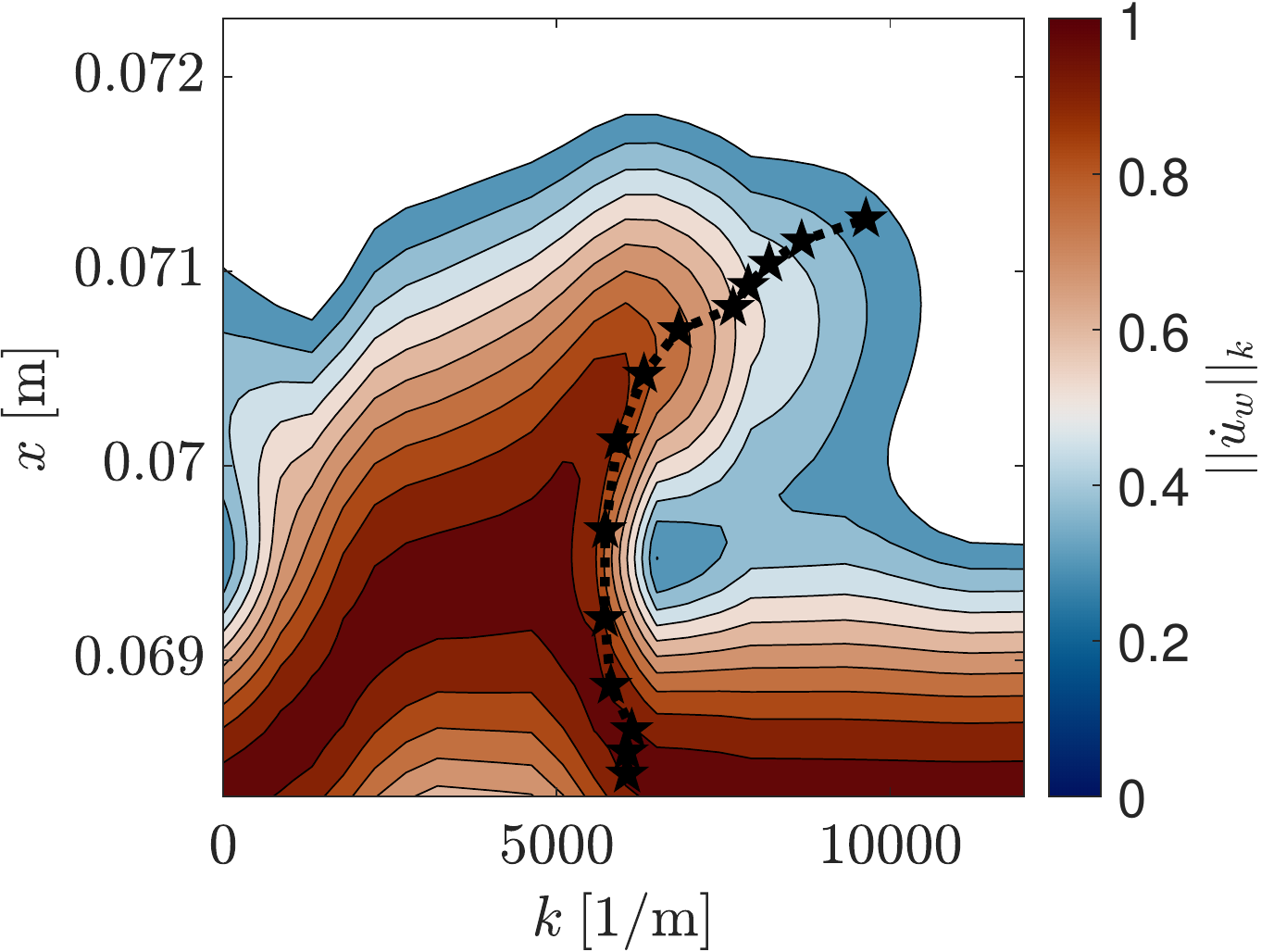}}\\
    \caption{Optimisation of the resonator spacing: (a) best RL discovered spacing when linear grading rules the resonator lengths; (b) spectrogram. In (a), $\zeta$--IPs are denoted with cyan markers. In (b), black markers correspond to the predictions of Eq. \ref{eq:DispersionRelation} for varying spacing and ratio $K_r/M_r$ for the resonators placed before the target resonator.\label{fig:8_3GeomSpectrogram}}
\end{figure}

\normalsize The best configuration discovered by the RL agent is reported on the top of Fig. \ref{fig:8_3GeomSpectrogram}. The number of resonators has been reduced from $35$ to $23$ by enlarging the spacing of the resonators placed before the target resonator. 
On the bottom, a spectrogram representation \cite{art:Riva20} has been employed to depict the outcome of the wave propagation obtained for this resonator arrangement.
The predictions of Eq. \ref{eq:DispersionRelation} agree well with the outcomes of the FE simulation even by simultaneously modifying the distance $D$ and the ratio $K_r/M_r$.

The ratio $R_{T}/R^H_{T}$ quantifying the performance of the optimisation procedure is equal to $3.600$. It implies just a $1.5\%$ performance improvement with respect to the case featuring a linear grading rule with $35$ resonators. This can can be addressed considering from Eq.\ref{eq:Skelton} that a change in height gives a more significant change in the wavevector with respect to a change in distance. Therefore, both analytical arguments and the outcomes of the optimisation procedure support the conclusion that playing on the resonator spacing is not as effective as playing on the resonator lengths for sake of mechanical energy confinement and amplification. However, this investigation has also revealed that employing the maximum allowed number of resonators is unnecessary for getting the best performance, therefore promising savings in the material and in the manufacturing costs.

\subsection{Optimisation by combining resonator lengths and spacing}
\label{sec:resultsLengthSpacing}

The joint optimisation of the resonator lengths and spacing has been tackled considering for the waveguide the same geometry of Sec. \ref{sec:resultsLengths}. The agent actions reported in Tab. \ref{tab:agentActionsLengthSpacing} have been used to modify the value of $N_s=10$ continuous variables by operating on: the coordinates of four $z$--IPs belonging to the $\left(x,z\right)$ plane and ruling the envelope $z=p\left(x\right)$ of the resonator lengths; the coordinates of four $\zeta$--IPs belonging to the $\left(x,\lambda \right)$ plane and ruling the envelope $\zeta=\tilde{p}\left(x\right)$ of the resonator spacing. The starting state of the continuous variable and the range of their possible values have been set similarly to what done in the previous sections. The number of neurons of the hidden layers used to approximate $d\left(s,a,\boldsymbol{\theta}_v\right)$ and to calibrate the policy parameters $\left[\mu\left(s,\boldsymbol{\theta}_p\right),\sigma\left(s,\boldsymbol{\theta}_p\right)\right]$, respectively, have been doubled to account for the larger dimension of the configuration space. The learning rate has been set to $1 \cdot 10^{-3}$ to speed up the training in view of the greater number of actions performed by the agent.

\begin{table}[h!]
\footnotesize
    \centering
    \[
    \begin{array}{cccc}
    \toprule
    \mbox{Action} & \mbox{What} & \mbox{Variable value} & \mbox{Range of} \\
    \mbox{ordering} & \mbox{is modified} & \mbox{starting state} & \mbox{possible values}\\
    \midrule
    1 & 1\mbox{st $z$--IP, }z & 5.028\cdot 10^{-4} & \left[0,9.156\cdot 10^{-4}\right]\mbox{ m} \\
    2 & 4\mbox{th $z$--IP, }z & 5.028\cdot 10^{-4} & \left[0,9.156\cdot 10^{-4}\right]\mbox{ m} \\
    3 & 2\mbox{nd $z$--IP, }x & 0.0698 \mbox{ m} & \left[0.0683,0.0713\right]\mbox{ m} \\
    4 & 2\mbox{nd $z$--IP, }z & 5.028\cdot 10^{-4} & \left[0,9.156\cdot 10^{-4} \right]\mbox{ m} \\
    5 & 1\mbox{st $\zeta$--IP, }\zeta & \lambda/16 & \left[0,\lambda/2\right] \\
    6 & 4\mbox{th $\zeta$--IP, }\zeta & \lambda/16 & \left[0,\lambda/2\right] \\
    7 & 2\mbox{nd $\zeta$--IP, }x & 0.0696 \mbox{ m} & \left[0.0683,0.0703\right]\mbox{ m} \\
    8 & 2\mbox{nd $\zeta$--IP, }\zeta & \lambda/16 & \left[0,\lambda/2\right] \\
    9 & 3\mbox{rd $\zeta$--IP, }x & 0.0710 & \left[0.0703,0.0723\right] \mbox{ m}\\
    10 & 3\mbox{rd $\zeta$--IP, }\zeta & \lambda/16 & \left[0,\lambda/2\right] \\
    \bottomrule
    \end{array}
    \]
    \caption{Optimisation of the resonator lengths and spacing: description of the agent actions. The $\zeta$--IPs are reported with cyan markers in Fig.; the $z$--IPs with orange markers in Fig..}
    \label{tab:agentActionsLengthSpacing}
\end{table}

\begin{figure*}
\centering
\subfloat[]{\label{fig:6_9geom}\includegraphics[width=0.45\textwidth]{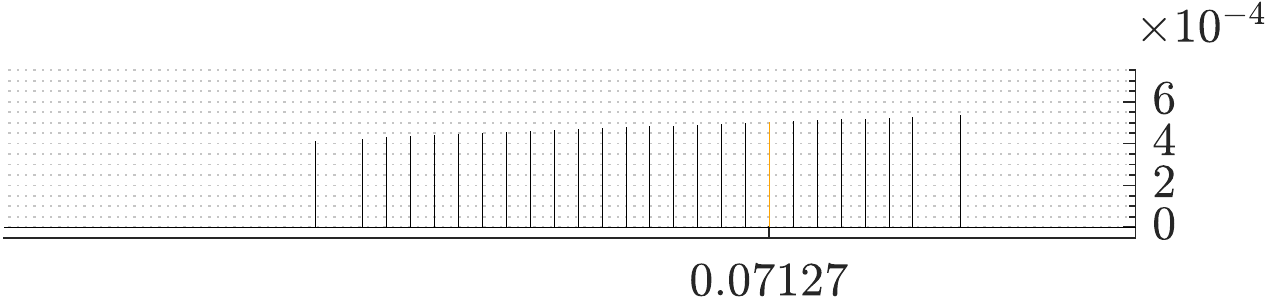}}
\subfloat[]{\label{fig:6_9_8_4_energyPlot}\includegraphics[width=0.4\textwidth]{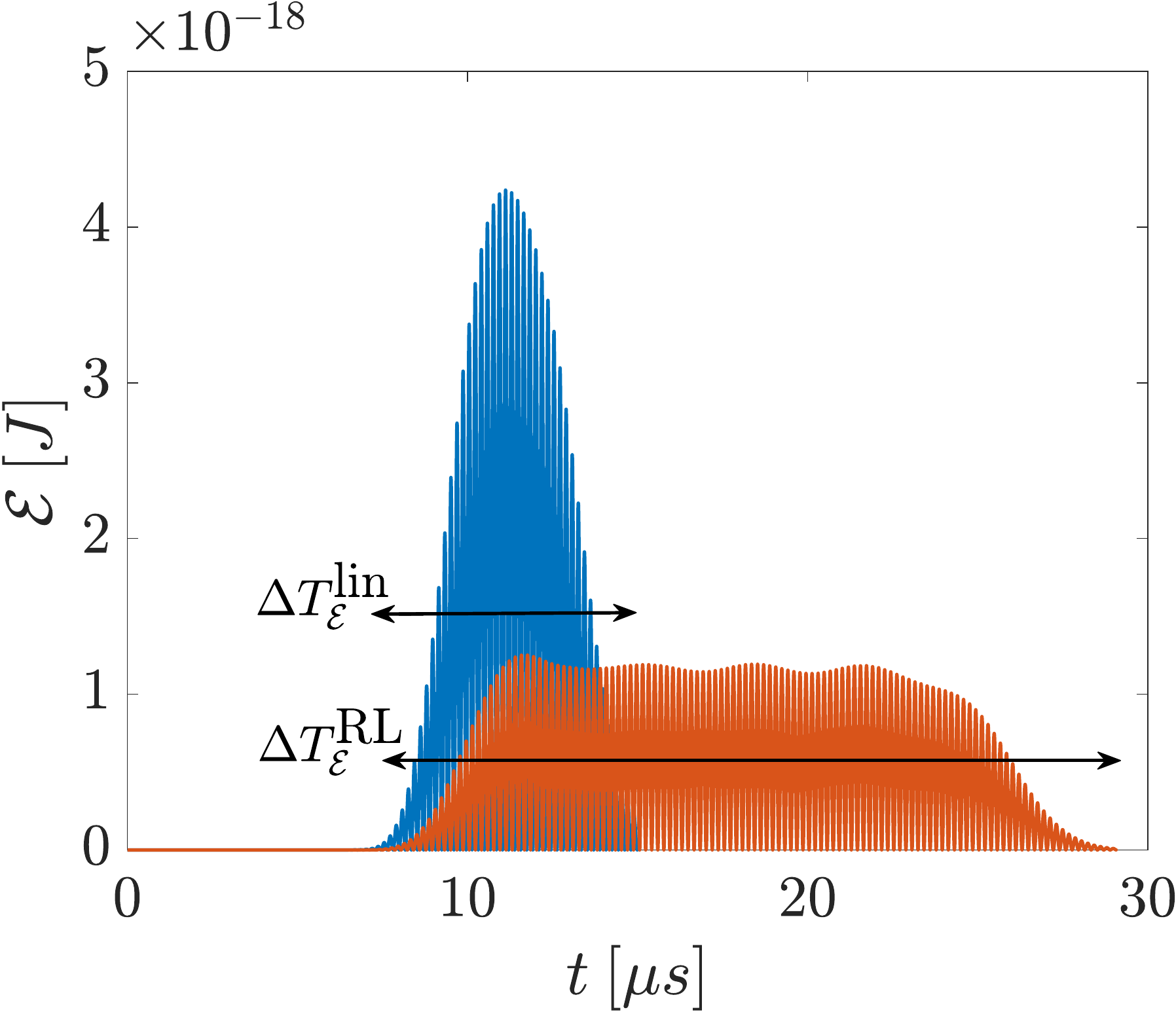}}$~$
\subfloat[]{\label{fig:WaterFall6_9_8_2}\includegraphics[width=0.55\textwidth]{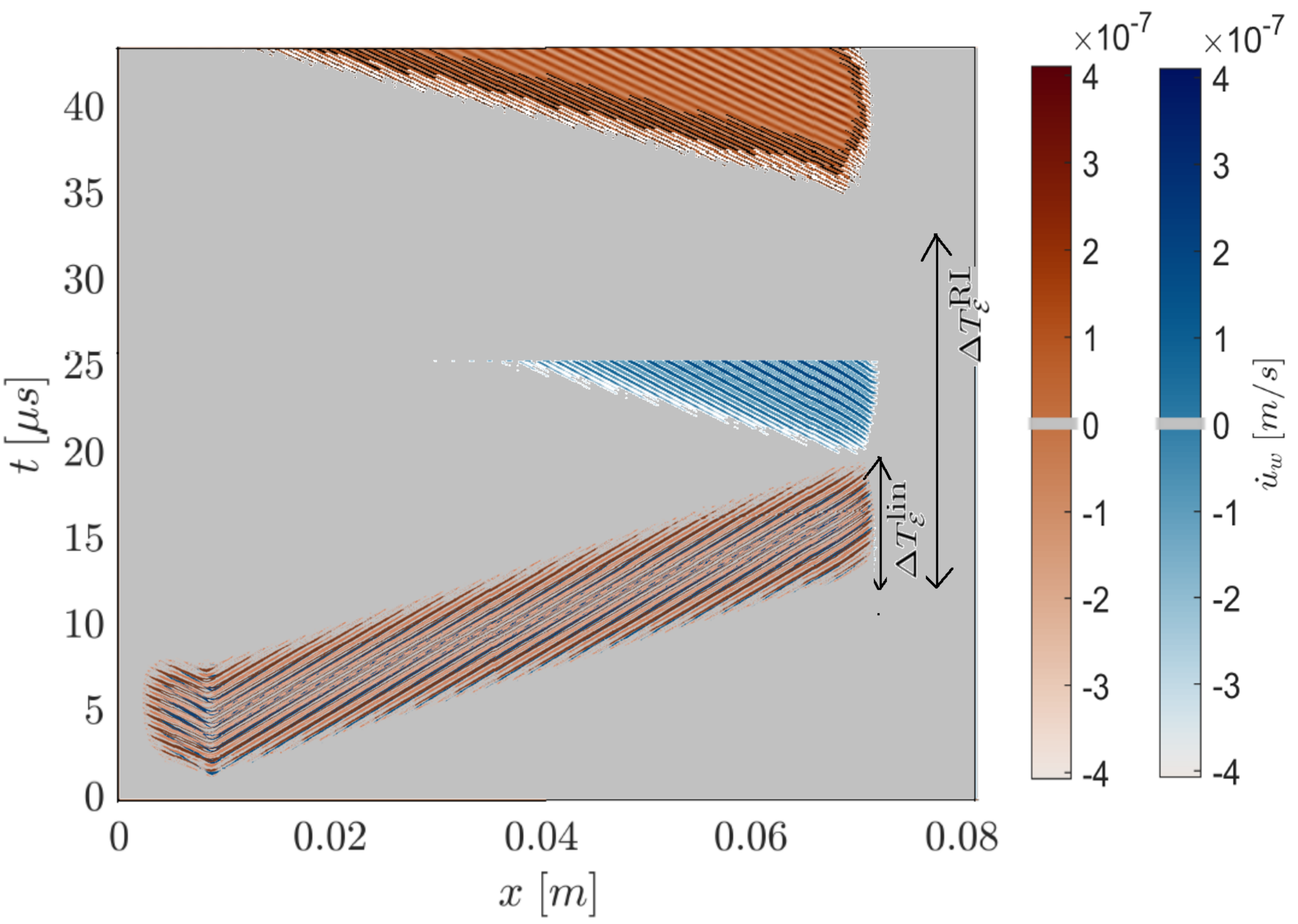}}\\
\caption{Optimisation of the resonator lengths and spacing. (a) Resonator arrangement for the optimised configuration (adaptive $T$ criterion). (b) Evolution of $\mathcal{E}$ during the analysis. In blue: evolution of $\mathcal{E}$ for the reference configuration \color{blue}{$R_{T}/R^H_{T}= 3.870$}. \color{black}{In} red: evolution of $\mathcal{E}$ for the best RL discovered configuration. \color{red}{$R_{T}/R^H_{T}= 4.365$}\color{black}{.} (c) Wave propagation time--spatial outcomes. Blue scale: outcomes for the reference configuration (see Fig. \ref{fig:8_2geom}. Red scale: outcomes for the best RL discovered configuration.}
\label{fig:optLengthsSpacingResults}
\end{figure*}

By running the optimisation, we notice that the outcomes strongly depend on the analysis time $T$ (see APPENDIX B).
It must be said that previously reported outcomes have not been affected by the adopted $T$, as in Sec. \ref{sec:resultsLengths} $T$ was set to a sufficiently large value, while the effect of the resonator spacing on the harvesting capabilities of the system has been shown to be limited in any case. However, the need of handling the possibly different time interval necessary to damp out the oscillations of the target resonator, and the possibly different interaction time between the propagating wave and the resonators justifies the need of employing an adaptive criterion to set $T$. On light of that, the next analyses have been stopped only when the $\mathcal{E}$ has been lower than $5\cdot 10^{-20} \mbox{J}$ for at least $500$ times in row. The count has been prescribed to start when the threshold value has been overcome for the first time.
The outcome of the optimisation procedure for an adaptive $T$ is reported in Fig. \ref{fig:6_9geom}. As in Sec. \ref{sec:resultsLengths}, the lengths of the resonators preceding the target resonator monotonically increases and features a negative concavity $p''\left(x\right)$. However, the slope of the grading is much lower than the one depicted in Fig. \ref{fig:bestConfigurationLengths}. Moreover, working on the spacing has enabled to take out some resonators from the guide. Despite the limited discrepancies, the mechanism exploited by the obtained configuration to increase $\mathcal{E}$ during analysis is largely different from the one featuring the linear grading of Fig. \ref{fig:8_2geom}. This reference resonator arrangement increases the time interval $\Delta T^{\text{lin}}_{\mathcal{E}}$ in which the target resonator oscillates with respect to a configuration featuring the target resonator without other resonators, but primarily focuses on amplifying the target resonator oscillations. On the other hand, the obtained resonator array works almost uniquely enlarging the oscillation time interval $\Delta T^{\text{RL}}_{\mathcal{E}}$, so that $\Delta T^{\text{RL}}_{\mathcal{E}}>\Delta T^{\text{lin}}_{\mathcal{E}}$ as clearly shown in Fig. \ref{fig:6_9_8_4_energyPlot} finally leading to enhance by $12.8\%$ the elastic energy exploitable. This increase has been possible despite target resonator oscillations even smaller than the ones related to the configuration featuring just the target resonator without other resonators. The time--spatial representation of \ref{fig:WaterFall6_9_8_2} further elucidates the matter by depicting how the wave packet propagates in the waveguide for the two cases. For both configurations, the wave travelling to the left has been immediately damped out by the absorbing conditions, while the wave propagating to the right has been converted in resonator oscillations. After a while, resonator oscillations have turned back to a wave packet left propagating in the waveguide. The oscillation conversion has taken place after $\Delta T^{\text{lin}}_{\mathcal{E}}$ for the blue curve depicting the wave propagation for the linear grading configuration, and after $\Delta T^{\text{RL}}_{\mathcal{E}}$ for the red curve referring to the optimised grading. The inequality $\Delta T^{\text{RL}}_{\mathcal{E}}>\Delta T^{\text{lin}}_{\mathcal{E}}$ has been allowed by the repeated wave reflections between the target resonator and the neighbouring resonators, as highlighted by the several $\mathcal{E}$ picks reported in red in Fig. \ref{fig:6_9_8_4_energyPlot}.

\section{Conclusions}
\label{sec:conclusions}
A reinforcement learning approach to optimise graded metamaterials for mechanical energy confinement and amplification has been proposed. The procedure leverages on constraining the design space on the basis of the knowledge of the physical behaviour of the system, and on the use of a policy gradient actor--critic algorithm to train the reinforcement learning algorithm. Results related to the optimisation of the grading rule of the resonator lengths and spacing highlight the potentiality of the method. The procedure can be applied to other design optimisation problems in which the effect of a design choice can be numerically simulated. A preliminary analytical investigation has been carried out by obtaining the dispersion relation for a local resonant metamaterial considering both variations in the resonator oscillating frequencies and spacing. The predictions of this dispersion relation have been found in agreement with numerical outcomes.

Specifically, results on the resonator lengths optimisation have confirmed previous works reported in the literature. The procedure has also highlighted the robustness for mechanical energy confinement and amplification of this type of mechanical system with respect to limited changes in the resonator configuration. Results on the resonator spacing have demonstrated the secondary effect of this design parameter with respect to the resonator lengths, however showing the possibility of reducing the number of resonator in the system still obtaining comparable, if not better, performances. Finally, the outcomes of the optimisation procedure when applied to the joint calibration of the resonator lengths and spacing with an adaptive rule to set the analysis duration has further highlighted the potentiality of the method with respect to a reference optimised configuration featuring a linear grading. The obtained configuration has focused almost uniquely on enlarging the time in which the target resonator oscillations take place without amplifying them. These outcomes have confirmed the fundamental importance that the analysis duration plays on the optimisation outcomes.

In future works, results for mechanical energy confinement and amplification will be used for energy harvesting in graded metamaterials exploiting the piezoelectric effect as transduction mechanism, possibly repeating the optimisation procedure for a modified reward related, for example, to the output voltage or power. Attention will be also devoted to design optimisation problems involving sources of uncertainty, for example the one related to the frequency content of the wave excitation, as well as the possibility of harvest energy on the entire set of resonators.

\section*{Acknowledgments}
The authors are grateful to Giacomo Arcieri (ETH Zurich) for fruitful discussions about reinforcement learning.
The support of the H2020 FET-proactive Metamaterial Enabled Vibration Energy Harvesting (MetaVEH) project under Grant Agreement No. 952039 is acknowledged.

\bibliographystyle{elsarticle-num}
\footnotesize
\bibliography{references}

\begin{thebibliography}{10}
\expandafter\ifx\csname url\endcsname\relax
  \def\url#1{\texttt{#1}}\fi
\expandafter\ifx\csname urlprefix\endcsname\relax\def\urlprefix{URL }\fi
\expandafter\ifx\csname href\endcsname\relax
  \def\href#1#2{#2} \def\path#1{#1}\fi

\bibitem{art:Smith2004}
D.~R. Smith, J.~B. Pendry, M.~C.~K. Wiltshire, Metamaterials and negative
  refractive index, Science 305~(5685) (2004) 788--792.
\newblock \href {https://doi.org/10.1126/science.1096796}
  {\path{doi:10.1126/science.1096796}}.

\bibitem{art:Pendry1999}
J.~Pendry, A.~Holden, D.~Robbins, W.~Stewart, Magnetism from conductors and
  enhanced nonlinear phenomena, IEEE Transactions on Microwave Theory and
  Techniques 47 (1999) 2075--2084.
\newblock \href {https://doi.org/10.1109/22.798002}
  {\path{doi:10.1109/22.798002}}.

\bibitem{art:Pendry2000}
J.~B. Pendry, Negative refraction makes a perfect lens, Physical Review Letters
  85 (2000) 3966--3969.
\newblock \href {https://doi.org/10.1103/PhysRevLett.85.3966}
  {\path{doi:10.1103/PhysRevLett.85.3966}}.

\bibitem{art:Liu2000}
Z.~Liu, X.~Zhang, Y.~Mao, Y.~Y. Zhu, Z.~Yang, C.~T. Chan, P.~Shen, Locally
  resonant sonic materials, Science 289 (2000) 1734--1736.
\newblock \href {https://doi.org/10.1126/science.289.5485.1734}
  {\path{doi:10.1126/science.289.5485.1734}}.

\bibitem{book:Craster2013}
R.~V. Craster, S.~Guenneau, Acoustic Metamaterials, Springer Dordrecht, 2016.
\newblock \href {https://doi.org/10.1007/978-94-007-4813-2}
  {\path{doi:10.1007/978-94-007-4813-2}}.

\bibitem{book:Craster2017}
R.~V. Craster, S.~Guenneau, World Scientific Handbook of Metamaterials and
  Plasmonics: Volume 2: Elastic, Acoustic and Seismic Metamaterials, Vol.~2,
  Singapore: World Scientific, 2017.

\bibitem{art:Moleron2015}
M.~Molerón, C.~Daraio, Acoustic metamaterial for subwavelength edge detection,
  Nature Communications 6~(8037) (2015) 1--6.
\newblock \href {https://doi.org/10.1038/ncomms9037}
  {\path{doi:10.1038/ncomms9037}}.

\bibitem{art:Ali2021}
M.~Syed Akbar~Ali, P.~Rajagopal, The promise of metamaterials for ultrasonic
  nondestructive evaluation, Advances in Non-destructive Evaluation (2021)
  381--–394\href {https://doi.org/10.1007/978-981-16-0186-6\_36}
  {\path{doi:10.1007/978-981-16-0186-6\_36}}.

\bibitem{book:Laude2015}
V.~Laude, Phononic Crystals Artificial Crystals for Sonic, Acoustic, and
  Elastic Waves, De Gruyter, 2015.
\newblock \href {https://doi.org/10.1515/9783110302660}
  {\path{doi:10.1515/9783110302660}}.

\bibitem{Matlack2016}
K.~H. Matlack, A.~Bauhofer, S.~Krödel, A.~Palermo, C.~Daraio, Composite
  3d-printed metastructures for low-frequency and broadband vibration
  absorption, Proceedings of the National Academy of Sciences 113~(30) (2016)
  8386--8390.
\newblock \href {https://doi.org/10.1073/pnas.1600171113}
  {\path{doi:10.1073/pnas.1600171113}}.

\bibitem{Brule2014}
S.~Br\^ul\'e, E.~H. Javelaud, S.~Enoch, S.~Guenneau, Experiments on seismic
  metamaterials: Molding surface waves, Physical Review Letters 112 (2014)
  133901.
\newblock \href {https://doi.org/10.1103/PhysRevLett.112.133901}
  {\path{doi:10.1103/PhysRevLett.112.133901}}.

\bibitem{Miniaci2016}
M.~Marco, K.~Anastasiia, B.~Federico, P.~Nicola~M., Large scale mechanical
  metamaterials as seismic shields, New Journal of Physics 18 (2016) 1--15.
\newblock \href {https://doi.org/10.1088/1367-2630/18/8/083041}
  {\path{doi:10.1088/1367-2630/18/8/083041}}.

\bibitem{Brule2020}
S.~Br\^ul\'e, S.~Enoch, S.~Guenneau, Emergence of seismic metamaterials:
  Current state and future perspectives, Physics Letters A 384~(1) (2020)
  126034.
\newblock \href {https://doi.org/10.1016/j.physleta.2019.126034}
  {\path{doi:10.1016/j.physleta.2019.126034}}.

\bibitem{Farhat2009}
M.~Farhat, S.~Guenneau, S.~Enoch, Ultrabroadband elastic cloaking in thin
  plates, Physical Review Letters 103 (2009) 024301.
\newblock \href {https://doi.org/10.1103/PhysRevLett.103.024301}
  {\path{doi:10.1103/PhysRevLett.103.024301}}.

\bibitem{Stenger2012}
N.~Stenger, M.~Wilhelm, M.~Wegener, Experiments on elastic cloaking in thin
  plates, Physical Review Letters 108 (2012) 014301.
\newblock \href {https://doi.org/10.1103/PhysRevLett.108.014301}
  {\path{doi:10.1103/PhysRevLett.108.014301}}.

\bibitem{Quadrelli2021}
D.~E. Quadrelli, R.~Craster, M.~Kadic, F.~Braghin, Elastic wave near-cloaking,
  Extreme Mechanics Letters 44 (2021) 101262.
\newblock \href {https://doi.org/10.1016/j.eml.2021.101262}
  {\path{doi:10.1016/j.eml.2021.101262}}.

\bibitem{KRUSHYNSKA2014179}
A.~Krushynska, V.~Kouznetsova, M.~Geers, Towards optimal design of locally
  resonant acoustic metamaterials, Journal of the Mechanics and Physics of
  Solids 71 (2014) 179--196.
\newblock \href {https://doi.org/10.1016/j.jmps.2014.07.004}
  {\path{doi:10.1016/j.jmps.2014.07.004}}.

\bibitem{MATOUS2017192}
K.~Matouš, M.~G. Geers, V.~G. Kouznetsova, A.~Gillman, A review of predictive
  nonlinear theories for multiscale modeling of heterogeneous materials,
  Journal of Computational Physics 330 (2017) 192--220.
\newblock \href {https://doi.org/10.1016/j.jcp.2016.10.070}
  {\path{doi:10.1016/j.jcp.2016.10.070}}.

\bibitem{Sridhar2016}
A.~Sridhar, V.~G. Kouznetsova, M.~G.~D. Geers, Homogenization of locally
  resonant acoustic metamaterials towards an emergent enriched continuum,
  Computational Mechanics 57 (2016) 423–--435.
\newblock \href {https://doi.org/10.1007/s00466-015-1254-y}
  {\path{doi:10.1007/s00466-015-1254-y}}.

\bibitem{LIU2021114161}
L.~Liu, A.~Sridhar, M.~Geers, V.~Kouznetsova, Computational homogenization of
  locally resonant acoustic metamaterial panels towards enriched continuum
  beam/shell structures, Computer Methods in Applied Mechanics and Engineering
  387 (2021) 114161.
\newblock \href {https://doi.org/10.1016/j.cma.2021.114161}
  {\path{doi:10.1016/j.cma.2021.114161}}.

\bibitem{Carrara2013}
M.~Carrara, M.~R. Cacan, J.~Toussaint, M.~J. Leamy, M.~Ruzzene, A.~Erturk,
  Metamaterial-inspired structures and concepts for elastoacoustic wave energy
  harvesting, Smart Materials and Structures 22 (2013) 1--12.
\newblock \href {https://doi.org/10.1088/0964-1726/22/6/065004}
  {\path{doi:10.1088/0964-1726/22/6/065004}}.

\bibitem{Mikoshiba2013}
K.~Mikoshiba, J.~M. Manimala, C.~Sun, Energy harvesting using an array of
  multifunctional resonators, Journal of Intelligent Material Systems and
  Structures 24~(2) (2013) 168--179.
\newblock \href {https://doi.org/10.1177/1045389X12460335}
  {\path{doi:10.1177/1045389X12460335}}.

\bibitem{Gonella2009}
S.~Gonella, A.~C. To, W.~K. Liu, Interplay between phononic bandgaps and
  piezoelectric microstructures for energy harvesting, Journal of the Mechanics
  and Physics of Solids 57~(3) (2009) 621--633.
\newblock \href {https://doi.org/10.1016/j.jmps.2008.11.002}
  {\path{doi:10.1016/j.jmps.2008.11.002}}.

\bibitem{Sugino2018}
C.~Sugino, A.~Erturk, Analysis of multifunctional piezoelectric metastructures
  for low-frequency bandgap formation and energy harvesting, Journal of Physics
  D: Applied Physics 51~(21) (2018) 1--12.
\newblock \href {https://doi.org/10.1088/1361-6463/aab97e}
  {\path{doi:10.1088/1361-6463/aab97e}}.

\bibitem{Chen2019}
J.-S. Chen, W.-J. Su, Y.~Cheng, W.-C. Li, C.-Y. Lin, A metamaterial structure
  capable of wave attenuation and concurrent energy harvesting, Journal of
  Intelligent Material Systems and Structures 30~(20) (2019) 2973--2981.
\newblock \href {https://doi.org/10.1177/1045389X19880023}
  {\path{doi:10.1177/1045389X19880023}}.

\bibitem{chp:DePonti2021}
J.~M. De~Ponti, Graded Elastic Metamaterials for Energy Harvesting, Springer
  International Publishing, Cham, 2021, pp. 27--60.
\newblock \href {https://doi.org/10.1007/978-3-030-69060-1\_3}
  {\path{doi:10.1007/978-3-030-69060-1\_3}}.

\bibitem{art:Tsakmakidis2007}
K.~Tsakmakidis, A.~Boardman, O.~Hess, ‘{Trapped} rainbow’ storage of light
  in metamaterials, Nature 450 (2007) 397–--401.

\bibitem{Zhu2013}
J.~Zhu, Y.~Chen, X.~Zhu, F.~J. Garcia-Vidal, X.~Yin, W.~Zhang, X.~Zhang,
  Acoustic rainbow trapping, Scientific Reports 3~(1728) (2013) 1--6.
\newblock \href {https://doi.org/10.1038/srep01728}
  {\path{doi:10.1038/srep01728}}.

\bibitem{Garcia2013}
V.~Romero-García, R.~Picó, A.~Cebrecos, V.~J. Sánchez-Morcillo,
  K.~Staliunas, Enhancement of sound in chirped sonic crystals, Applied Physics
  Letters 102~(091906) (2013) 1--5.
\newblock \href {https://doi.org/10.1063/1.4793575}
  {\path{doi:10.1063/1.4793575}}.

\bibitem{Garcia2014}
A.~Cebrecos, R.~Picó, V.~J. Sánchez-Morcillo, K.~Staliunas,
  V.~Romero-García, L.~M. Garcia-Raffi, Enhancement of sound by soft
  reflections in exponentially chirped crystals, AIP Advances 4~(124402) (2014)
  1--11.

\bibitem{art:Colombi2016}
A.~Colombi, D.~Colquitt, P.~Roux, S.~Guenneau, R.~V. Craster, A seismic
  metamaterial: The resonant metawedge, Scientific Reports 6~(27717) (2016)
  1--6.
\newblock \href {https://doi.org/10.1038/srep27717}
  {\path{doi:10.1038/srep27717}}.

\bibitem{Chaplain2020Delineating}
G.~J. Chaplain, D.~Pajer, J.~M.~D. Ponti, R.~V. Craster, Delineating rainbow
  reflection and trapping with applications for energy harvesting, New Journal
  of Physics 22~(063024) (2020) 1--13.
\newblock \href {https://doi.org/10.1088/1367-2630/ab8cae}
  {\path{doi:10.1088/1367-2630/ab8cae}}.

\bibitem{art:Chaplain2020Umklapp}
G.~J. Chaplain, J.~M.~D. Ponti, A.~Colombi, R.~Fuentes-Dominguez, P.~Dryburg,
  D.~Pieris, R.~J. Smith, A.~Clare, M.~Clark, R.~V. Craster, Tailored elastic
  surface to body wave umklapp conversion, Nature Communications 11~(3267)
  (2020) 1--6.
\newblock \href {https://doi.org/10.1038/s41467-020-17021-x}
  {\path{doi:10.1038/s41467-020-17021-x}}.

\bibitem{art:NJP20}
J.~M. De~Ponti, A.~Colombi, R.~Ardito, F.~Braghin, A.~Corigliano, R.~V.
  Craster, Graded elastic metasurface for enhanced energy harvesting, New
  Journal of Physics 22~(1) (2020) 013013.
\newblock \href {https://doi.org/10.1088/1367-2630/ab6062}
  {\path{doi:10.1088/1367-2630/ab6062}}.

\bibitem{art:APL20}
J.~M. De~Ponti, A.~Colombi, E.~Riva, R.~Ardito, F.~Braghin, A.~Corigliano,
  R.~V. Craster, Experimental investigation of amplification, via a mechanical
  delay--line, in a rainbow--based metamaterial for energy harvesting, Applied
  Physics Letters 117~(14) (2020) 143902.
\newblock \href {https://doi.org/10.1063/5.0023544}
  {\path{doi:10.1063/5.0023544}}.

\bibitem{Zhao2022}
B.~Zhao, H.~R. Thomsen, J.~M. {De Ponti}, E.~Riva, B.~{Van Damme},
  A.~Bergamini, E.~Chatzi, A.~Colombi, A graded metamaterial for broadband and
  high-capability piezoelectric energy harvesting, Energy Conversion and
  Management 269 (2022) 116056.
\newblock \href {https://doi.org/10.1016/j.enconman.2022.116056}
  {\path{doi:10.1016/j.enconman.2022.116056}}.

\bibitem{ErturkGrading2020}
M.~Alshaqaq, A.~Erturk, Graded multifunctional piezoelectric metastructures for
  wideband vibration attenuation and energy harvesting, Smart Materials and
  Structures 30~(015029) (2020) 1--11.
\newblock \href {https://doi.org/10.1088/1361-665x/abc7fa}
  {\path{doi:10.1088/1361-665x/abc7fa}}.

\bibitem{ErturkGrading2022}
M.~Alshaqaq, C.~Sugino, A.~Erturk, Programmable rainbow trapping and band-gap
  enhancement via spatial group-velocity tailoring in elastic metamaterials,
  Physical Review Applied 17 (2022) L021003.
\newblock \href {https://doi.org/10.1103/PhysRevApplied.17.L021003}
  {\path{doi:10.1103/PhysRevApplied.17.L021003}}.

\bibitem{ErturkGrading2022b}
J.~Santini, C.~Sugino, E.~Riva1, A.~Erturk, Harnessing rainbow trapping via
  hybrid electromechanical metastructures for enhanced energy harvesting and
  vibration attenuation, Journal of Applied Physics 132~(064903) (2022) 1--15.
\newblock \href {https://doi.org/10.1063/5.0090258}
  {\path{doi:10.1063/5.0090258}}.

\bibitem{JianGrading2022}
Y.~Jian, L.~Tang, G.~Hu, Z.~Li, K.~C. Aw, Design of graded piezoelectric
  metamaterial beam with spatial variation of electrodes, International Journal
  of Mechanical Sciences 218 (2022) 107068.
\newblock \href {https://doi.org/10.1016/j.ijmecsci.2022.107068}
  {\path{doi:10.1016/j.ijmecsci.2022.107068}}.

\bibitem{art:Ororbia21}
M.~E. Ororbia, G.~P. Warn, {Design Synthesis Through a Markov Decision Process
  and Reinforcement Learning Framework}, Journal of Computing and Information
  Science in Engineering 22~(2), 021002 (07 2021).
\newblock \href {https://doi.org/10.1115/1.4051598}
  {\path{doi:10.1115/1.4051598}}.

\bibitem{book:Sutton18}
R.~S. Sutton, A.~G. Barto, Reinforcement learning: An introduction, MIT press,
  Cambridge, MA, 2018.

\bibitem{SKINNER2018933}
S.~Skinner, H.~Zare-Behtash, State-of-the-art in aerodynamic shape optimisation
  methods, Applied Soft Computing 62 (2018) 933--962.
\newblock \href {https://doi.org/10.1016/j.asoc.2017.09.030}
  {\path{doi:10.1016/j.asoc.2017.09.030}}.

\bibitem{art:Jenkins91}
W.~Jenkins, Towards structural optimization via the genetic algorithm,
  Computers \& Structures 40~(5) (1991) 1321--1327, special Issue:
  Computational Structures Technology.
\newblock \href {https://doi.org/10.1016/0045-7949(91)90402-8}
  {\path{doi:10.1016/0045-7949(91)90402-8}}.

\bibitem{art:Perez07}
R.~Perez, K.~Behdinan, Particle swarm approach for structural design
  optimization, Computers \& Structures 85~(19) (2007) 1579--1588.
\newblock \href {https://doi.org/10.1016/j.compstruc.2006.10.013}
  {\path{doi:10.1016/j.compstruc.2006.10.013}}.

\bibitem{art:Viquerat21}
J.~Viquerat, J.~Rabault, A.~Kuhnle, H.~Ghraieb, A.~Larcher, E.~Hachem, Direct
  shape optimization through deep reinforcement learning, Journal of
  Computational Physics 428 (2021) 110080.
\newblock \href {https://doi.org/10.1016/j.jcp.2020.110080}
  {\path{doi:10.1016/j.jcp.2020.110080}}.

\bibitem{art:Fan20}
D.~Fan, L.~Yang, Z.~Wang, M.~S. Triantafyllou, G.~E. Karniadakis, Reinforcement
  learning for bluff body active flow control in experiments and simulations,
  Proceedings of the National Academy of Sciences 117~(42) (2020) 26091--26098.
\newblock \href {https://doi.org/10.1073/pnas.2004939117}
  {\path{doi:10.1073/pnas.2004939117}}.

\bibitem{art:Pahlavani22}
H.~Pahlavani, M.~Amani, M.~C. {Sald\'{i}var}, J.~Zhou, M.~M. J., Z.~A. A., Deep
  learning for the rare--event rational design of 3d printed multi--material
  mechanical metamaterials, Communications Materials 3~(46) (2022).
\newblock \href {https://doi.org/10.1038/s43246-022-00270-2}
  {\path{doi:10.1038/s43246-022-00270-2}}.

\bibitem{art:Papadrakakis98}
M.~Papadrakakis, N.~D. Lagaros, Y.~Tsompanakis, Structural optimization using
  evolution strategies and neural networks, Computer Methods in Applied
  Mechanics and Engineering 156~(1) (1998) 309--333.
\newblock \href {https://doi.org/10.1016/S0045-7825(97)00215-6}
  {\path{doi:10.1016/S0045-7825(97)00215-6}}.

\bibitem{art:Cagan05}
J.~Cagan, M.~I. Campbell, S.~Finger, T.~Tomiyama, A framework for computational
  design synthesis: Model and applications, Journal of Computing and
  Information Science in Engineering 5~(3) (2005) 171--181.
\newblock \href {https://doi.org/10.1115/1.2013289}
  {\path{doi:10.1115/1.2013289}}.

\bibitem{art:Greensmith04}
E.~Greensmith, P.~L. Bartlett, J.~Baxter, Variance reduction techniques for
  gradient estimates in reinforcement learning., Journal of Machine Learning
  Research 5~(9) (2004).

\bibitem{art:Barto83}
A.~G. Barto, R.~S. Sutton, C.~W. Anderson, Neuronlike adaptive elements that
  can solve difficult learning control problems, IEEE Transactions on Systems,
  Man, and Cybernetics SMC-13~(5) (1983) 834--846.
\newblock \href {https://doi.org/10.1109/TSMC.1983.6313077}
  {\path{doi:10.1109/TSMC.1983.6313077}}.

\bibitem{proc:Sutton99}
R.~S. Sutton, D.~McAllester, S.~Singh, Y.~Mansour, Policy gradient methods for
  reinforcement learning with function approximation, in: S.~Solla, T.~Leen,
  K.~M\"{u}ller (Eds.), Advances in Neural Information Processing Systems,
  Vol.~12, MIT Press, 1999.

\bibitem{art:Schulman17}
J.~Schulman, F.~Wolski, P.~Dhariwal, A.~Radford, O.~Klimov, Proximal policy
  optimization algorithms (2017).
\newblock \href {https://doi.org/10.48550/ARXIV.1707.06347}
  {\path{doi:10.48550/ARXIV.1707.06347}}.

\bibitem{art:Rumelhart86}
D.~E. Rumelhart, G.~E. Hinton, R.~J. Williams, Learning representations by
  back--propagating errors, Nature 323 (1986) 533--536.
\newblock \href {https://doi.org/10.1038/323533a0}
  {\path{doi:10.1038/323533a0}}.

\bibitem{proc:Kingma15}
D.~Kingma, J.~Ba, Adam: A method for stochastic optimization, San Diego, CA,
  2015, pp. 1--13.

\bibitem{proc:Schulman15}
J.~Schulman, S.~Levine, P.~Abbeel, M.~Jordan, P.~Moritz,
  \href{https://proceedings.mlr.press/v37/schulman15.html}{Trust region policy
  optimization}, in: F.~Bach, D.~Blei (Eds.), Proceedings of the 32nd
  International Conference on Machine Learning, Vol.~37 of Proceedings of
  Machine Learning Research, PMLR, Lille, France, 2015, pp. 1889--1897.
\newline\urlprefix\url{https://proceedings.mlr.press/v37/schulman15.html}

\bibitem{art:Rajagopal12}
P.~Rajagopal, M.~Drozdz, E.~A. Skelton, M.~J. Lowe, R.~V. Craster, On the use
  of absorbing layers to simulate the propagation of elastic waves in unbounded
  isotropic media using commercially available finite element packages, NDT \&
  E International 51 (2012) 30--40.
\newblock \href {https://doi.org/10.1016/j.ndteint.2012.04.001}
  {\path{doi:10.1016/j.ndteint.2012.04.001}}.

\bibitem{art:OBoy10}
D.~O’Boy, V.~Krylov, V.~Kralovic, Damping of flexural vibrations in
  rectangular plates using the acoustic black hole effect, Journal of Sound and
  Vibration 329~(22) (2010) 4672--4688.
\newblock \href {https://doi.org/10.1016/j.jsv.2010.05.019}
  {\path{doi:10.1016/j.jsv.2010.05.019}}.

\bibitem{art:Georgiev11}
V.~Georgiev, J.~Cuenca, F.~Gautier, L.~Simon, V.~Krylov, Damping of structural
  vibrations in beams and elliptical plates using the acoustic black hole
  effect, Journal of Sound and Vibration 330~(11) (2011) 2497--2508.
\newblock \href {https://doi.org/10.1016/j.jsv.2010.12.001}
  {\path{doi:10.1016/j.jsv.2010.12.001}}.

\bibitem{art:Skelton18}
E.~A. Skelton, R.~V. Craster, A.~Colombi, D.~J. Colquitt, The multi-physics
  metawedge: graded arrays on fluid-loaded elastic plates and the mechanical
  analogues of rainbow trapping and mode conversion, New Journal of Physics
  20~(5) (2018) 053017.
\newblock \href {https://doi.org/10.1088/1367-2630/aabecf}
  {\path{doi:10.1088/1367-2630/aabecf}}.

\bibitem{art:Lemoult11}
F.~Lemoult, M.~Fink, G.~Lerosey, Acoustic resonators for far-field control of
  sound on a subwavelength scale, Physical Review Letters 107 (2011) 064301.
\newblock \href {https://doi.org/10.1103/PhysRevLett.107.064301}
  {\path{doi:10.1103/PhysRevLett.107.064301}}.

\bibitem{book:Belytschko00}
T.~Belytschko, W.~Liu, B.~Moran, Nonlinear Finite Elements for Continua and
  Structures, John Wiley \& Sons, Ltd, Chichester, United Kingdom, 2000.

\bibitem{code:stableBaselines}
A.~Hill, A.~Raffin, M.~Ernestus, A.~Gleave, A.~Kanervisto, R.~Traore,
  P.~Dhariwal, C.~Hesse, O.~Klimov, A.~Nichol, M.~Plappert, A.~Radford,
  J.~Schulman, S.~Sidor, Y.~Wu, Stable baselines,
  \url{https://github.com/hill-a/stable-baselines} (2018).

\bibitem{art:Riva20}
E.~Riva, M.~I.~N. Rosa, M.~Ruzzene, Edge states and topological pumping in
  stiffness-modulated elastic plates, Physical Review B 101 (2020) 094307.
\newblock \href {https://doi.org/10.1103/PhysRevB.101.094307}
  {\path{doi:10.1103/PhysRevB.101.094307}}.

\end{thebibliography}

\newpage
\section*{APPENDIX A}

\normalsize We report the complete PPO algorithm employed to solve the MDP with which the design process is formalised. Rewards are computed through FE simulations. Additional details on PPO can be found in \cite{proc:Schulman15}. As it can be noted, a major element of interest of RL with respect to supervised learning is the possibility to collect data during the training, potentially online.

\begin{algorithm}
\begin{algorithmic}[1]
\State Initialise the actor parameters $\boldsymbol{\theta}_{p}$ randomly.
\State Initialise the critic parameters $\boldsymbol{\theta}_{v}$ randomly.
\State $\boldsymbol{\theta}_{p_{\text{old}}}=\boldsymbol{\theta}_{p}$, $n_i=0$.
\While{$n_i<N_i$}
\For{$n_e=1,\ldots,N_e$}
\State Initialise the starting state $S^{n_e}_0=S_0$,
\For{$n_t=1,\ldots,N_t$}
\State $n_i=n_i+1$.
\State Sample $A^{n_e}_{n_t}$ from $\pi_{\text{old}}\left(a|S^{n_e}_{n_t-1},\boldsymbol{\theta}_{p_{\text{old}}}\right)$.
\State Determine $S^{n_e}_{n_t}$.
\State Compute $R^{n_e}_{n_t}$ via FEs.
\EndFor
\State Compute $G^{n_e}_{n_t}$ as in Eq. \eqref{eq:expectedReturn} for $n_t=1,\ldots
,N_t$.
\EndFor
\State Compute $d\left(S^{n_e}_{n_t},A^{n_e}_{n_t},\boldsymbol{\theta}_{v}\right)$.
\State Compute  $\mathcal{L}_p\left(\boldsymbol{\theta}_{p}\right)$ as in Eq. \eqref{eq:PPOobjective} .
\State Update $\boldsymbol{\theta}_{p}$ via gradient ascend to maximise $\mathcal{L}_p\left(\boldsymbol{\theta}_{p}\right)$.
\State Compute $\mathcal{L}_v\left(\boldsymbol{\theta}_{v}\right)$ as in Eq. \eqref{eq:criticObjective}.
\State Update $\boldsymbol{\theta}_{v}$ via gradient descend to minimise $\mathcal{L}_v\left(\boldsymbol{\theta}_{v}\right)$.
\State Set $\boldsymbol{\theta}_{p_{\text{old}}}=\boldsymbol{\theta}_{p}$.
\EndWhile
\end{algorithmic}
\caption{MDP solution via PPO for rewards calculated through FEs.}
\label{al:PPOforMDP}
\end{algorithm}

\section*{APPENDIX B}

\normalsize We show the optimisation results for fixed analysis time. A first optimisation attempt in which the analysis time $T$ has been set to $1.5\cdot 10^{-5}$ has led to contradictory outcomes highlighting an unusual decreasing profile of the lengths of the resonators preceding the target resonator, as shown in Fig. \ref{fig:6_6geom}.

\setcounter{figure}{0} 
\renewcommand{\figurename}{Figure B.}

\begin{figure}
\centering
\subfloat[Best RL discovered configuration for $T=1.5\cdot 10^{-5}$ s.]{\label{fig:6_6geom}\includegraphics[width=0.45\textwidth]{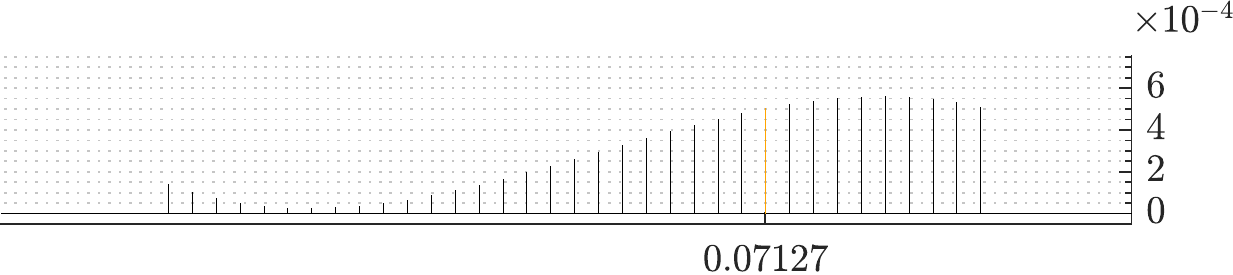}}
\quad \subfloat[$R_{T}/R^H_{T}= 3.052$, \color{orange}{$R_{T^*}/R^H_{T^*}= 3.566$.}]{\label{fig:6_6_6_7_energyPlot}\includegraphics[width=0.45\textwidth]{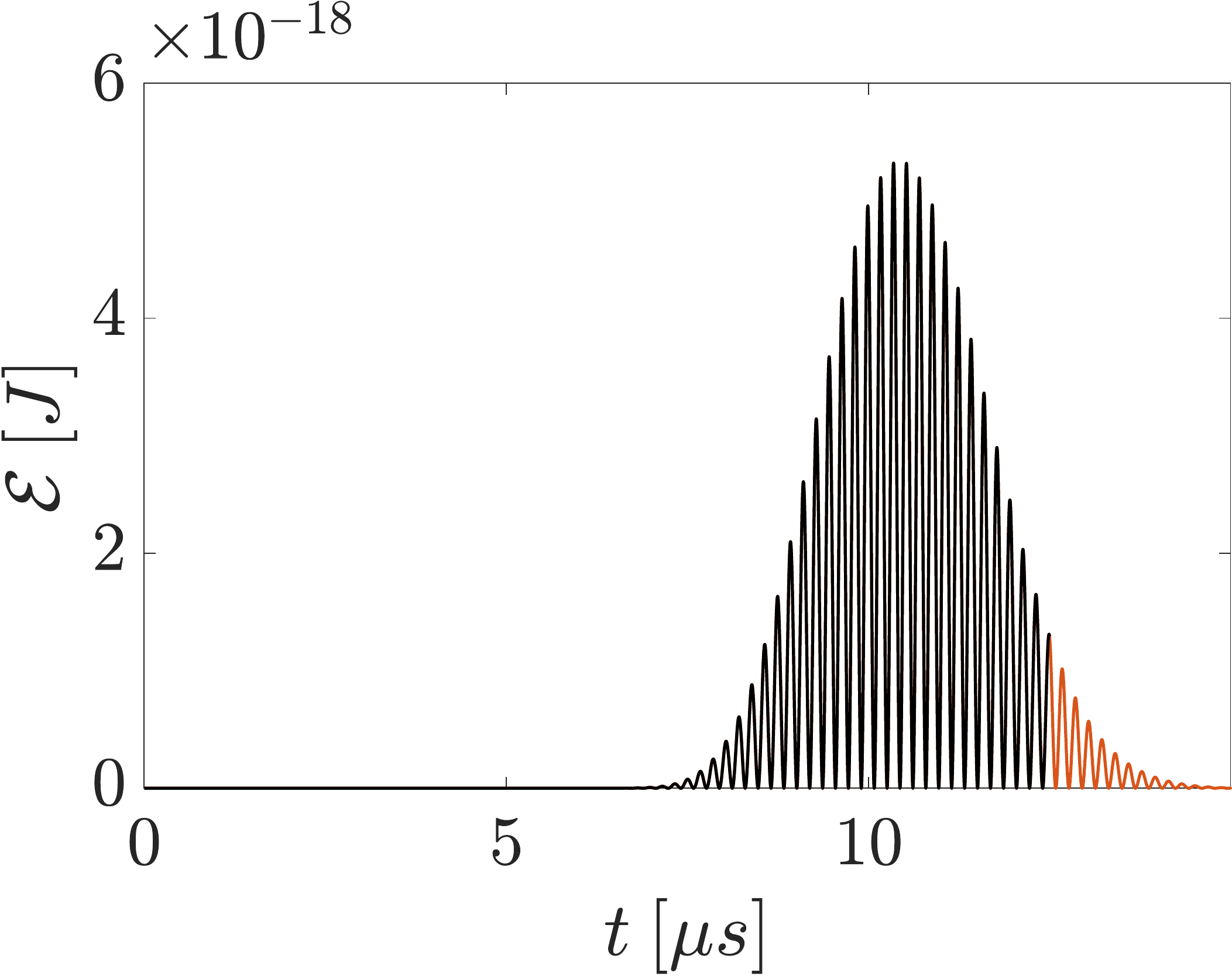}}
\\
\subfloat[Reference configuration (linear grading, spacing fixed to $\lambda/16$).]{\label{fig:8_2geom}\includegraphics[width=0.45\textwidth]{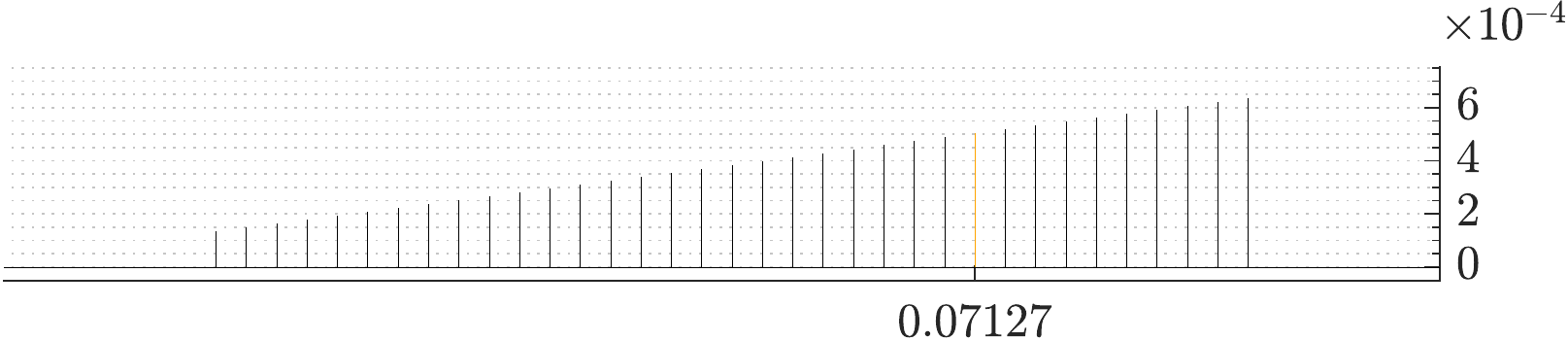}}
\quad \subfloat[$R_{T}/R^H_{T}= 2.669$, \color{orange}{$R_{T^*}/R^H_{T^*}= 3.870$.}]{\label{fig:8_2_8_4_energyPlot}\includegraphics[width=0.45\textwidth]{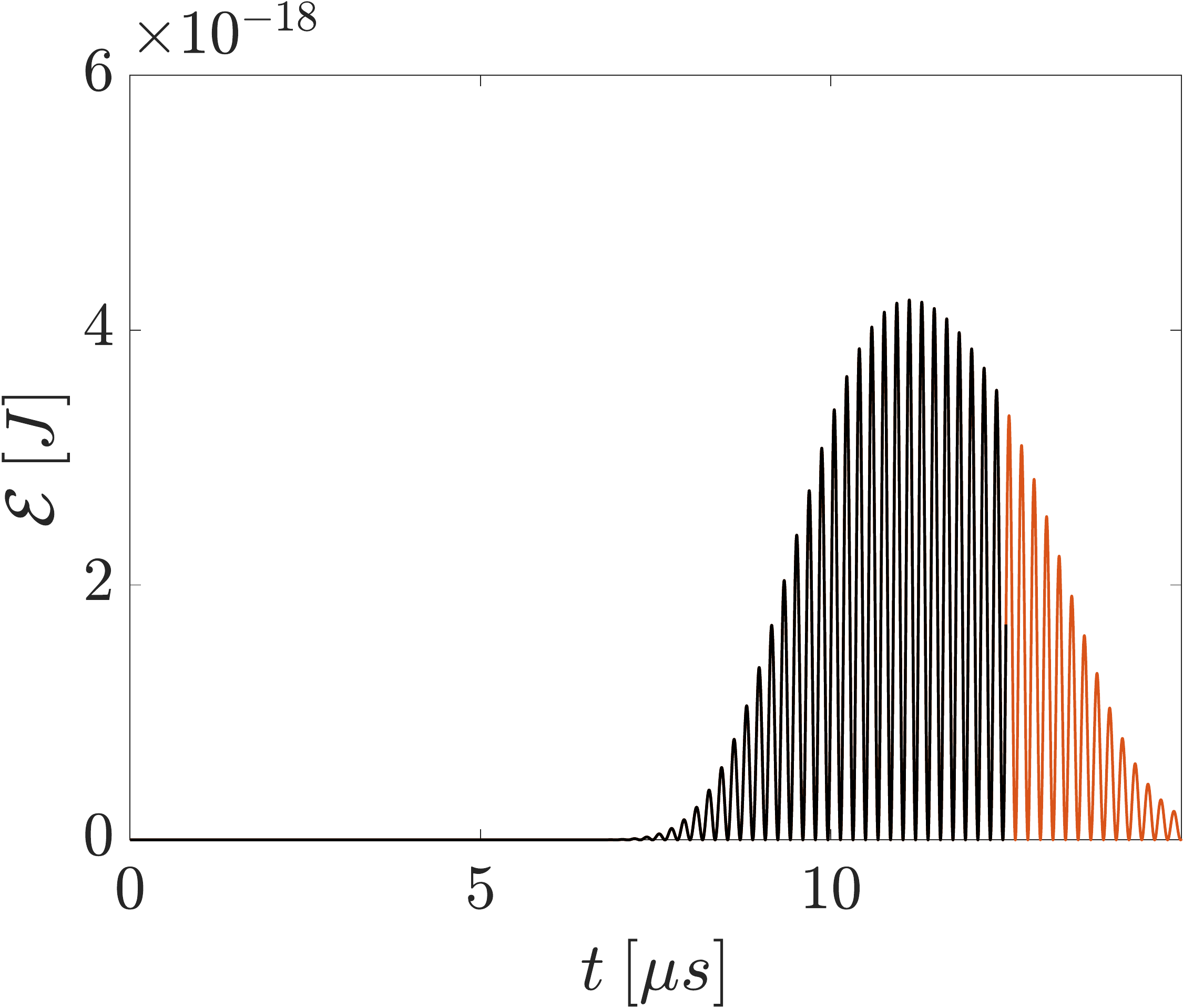}}
\caption{Optimisation of the resonator lengths and spacing: effect of the analysis time T on the optimisation outcomes. In (b) and (d), the time evolution of $\mathcal{E}$ for $T=1.5\cdot 10^{-5}$ (in black) are compared with those obtained for $T^*=2.0\cdot 10^{-5}$ (in red). (a) reports the resonator arrangement obtained by the optimisation procedure considering $T$ as duration of the wave propagation.}
\label{fig:effectOfT}
\end{figure}

This result disagrees both with what reported in Sec. \ref{sec:resultsLengths} and with the literature, see \cite{art:NJP20,art:APL20}. It has been necessary to inspect the evolution of the target resonator $\mathcal{E}$ to explain the reason behind this unexpected outcome. As shown in Fig. B\ref{fig:8_2_8_4_energyPlot}, the analysis has been too short to damp out the oscillations of the target resonator when the linear grading rule of Fig. B\ref{fig:8_2geom} has been employed. Consequently, the agent discovered a resonator configuration capable of exciting stronger but for a shorter time the target resonator. Computing the ratio $R_{T^*}/R^H_{T^*}$ for $T=2.0\cdot 10^{-5}$ for the obtained configuration and for a resonator arrangement ruled by a linear grading confirms what has been suggested, highlighting the better performance of the linear grading rule.

\end{document}